\documentclass[a4paper,usenatbib,useAMS]{mnras}
\usepackage{graphicx}
\usepackage{lscape}
\usepackage{perpage}
\usepackage{amsmath}
\usepackage{color}
\usepackage[normalem]{ulem}
\MakePerPage{footnote}

\newcommand{\beq}	{\begin{equation}}
\newcommand{\eeq}	{\end{equation}}
\newcommand{\beqa}	{\begin{eqnarray}}
\newcommand{\eeqa}	{\end{eqnarray}}
\newcommand{\e}	        {$^{-1}$}
\newcommand{\ee}	{$^{-2}$}
\newcommand{\eee}	{$^{-3}$}
\newcommand{\dis}		{\displaystyle}

\newcommand{\calm}	{{\cal M}}

\newcommand{\avg}[1]    {{\langle #1 \rangle}} 

\newcommand{\vecB}	{{\bf B}}

\newcommand{\vecv}	{{\bf v}}
\newcommand{\vecx}	{{\bf x}}

\newcommand\fnt		{\footnotetext}

\newcommand{\alfven}    {{Alfv$\acute{\rm e}$n }}

\newcommand{\alfvenic}  {{Alfv$\acute{\rm e}$nic }}
\newcommand{\alfvenicstop}  {{Alfv$\acute{\rm e}$nic}}

\newcommand{\muphi}	{\mu_{\Phi}}

\newcommand{\va}	{v_{\rm A}}

\newcommand{\avir}      {\alpha_{\rm vir,f}}

\newcommand{\ma}	{{\calm_{\rm A}}}

\newcommand{\mug}	{$\mu$G}
\newcommand{\mmug}	{\mu{\rm G}}

\newcommand{\rms}       {{\rm rms}}

\newcommand{\um}    {$\mu$m}

\newcommand\cs		{c_{\rm s}}

\newcommand{\htwom}      {{\rm H_2}}

\newcommand\POS         {{\rm POS}}

\newcommand{\tot}		{{\rm tot}}

\newcommand{\bx}		{\vecB(\vecx)}
\newcommand{\bxl}	    {\vecB(\vecx+\vecl)}
\newcommand{\boox}		{\vecB_0(\vecx)}
\newcommand{\booxl}		{\vecB_0(\vecx+\vecl)}

\newcommand{\botd}      {B_{\rm 0,\,3D}}
\newcommand{\btd}      {B_{\rm 3D}}
\newcommand{\crit}		{{\rm crit}}
\newcommand{\degree}	    {{$^\circ$}}
\newcommand{\fdcf}      {f_{\rm DCF}}
\newcommand{\muphipos}		{\mu_{\Phi,\POS}}
\newcommand{\mbe}       {M_{\rm BE}}
\newcommand{\mpl}		{M_{\Phi,\ell}}
\newcommand{\sdb}		{\sigma_{\delta B_\perp}}
\newcommand{\tfa}       {{\theta_{\rm FA}}}
\newcommand{\tfai}      {\theta_{{\rm FA},i}}
\newcommand{\vecl}      {\boldsymbol{\ell}}
\newcommand{\vir}		{{\rm vir}}

\usepackage{color}

\title[Magnetic field in Taurus/B211]{Mapping the magnetic field in the Taurus/B211 filamentary cloud with SOFIA HAWC+ and comparing with simulation}
\author[Pak Shing Li et al.]{
Pak Shing Li$^{1}$\thanks{E-mail:psli@berkeley.edu} (PSL),
Enrique Lopez-Rodriguez$^{2}$, 
Hamza Ajeddig$^{3}$,
\newauthor
~Philippe Andr\'e$^{3}$,
Christopher F. McKee$^{1}$,
Jeonghee Rho$^{4}$, 
Richard I. Klein$^{1,5}$
\\
$^{1}$Astronomy Department, University of California, Berkeley, CA 94720\\
$^{2}$Kavli Institute for Particle Astrophysics and Cosmology (KIPAC), Stanford University, Stanford, CA 94305, USA \\
$^{3}$ Laboratoire d'Astrophysique (AIM), CEA/DRF, CNRS, Universit\'e Paris-Saclay, Universit\'e Paris Diderot, \\
Sorbonne Paris Cit\'e, 91191 Gif-sur-Yvette, France \\
$^{4}$SETI Institute, 189 N. Bernardo Ave., Ste. 200, Mountain View, CA 94043 \\
$^{5}$Lawrence Livermore National Laboratory,P.O.Box 808, L-23, Livermore, CA 94550\\
}

\begin{document}

\date{}

\pubyear{}

\label{firstpage}
\pagerange{\pageref{firstpage}--\pageref{lastpage}}
\maketitle

\begin{abstract}

Optical and infrared polarization mapping and recent Planck observations of the filametary cloud L1495 in Taurus show that the large-scale magnetic field is approximately perpendicular to the long axis of the cloud.  We use the HAWC+ polarimeter on SOFIA to probe the complex magnetic field in the B211 part of the cloud. Our results reveal a dispersion of polarization angles of $36^\circ$, about five times that measured on a larger scale by Planck. Applying the Davis-Chandrasekhar-Fermi (DCF) method with velocity information obtained from IRAM 30m C$^{18}$O(1-0) observations, we find two distinct sub-regions with magnetic field strengths differing by more than a factor 3. The quieter sub-region is magnetically critical and sub-\alfvenicstop; the field is comparable to the average field measured in molecular clumps based on Zeeman observations. The more chaotic, super-\alfvenic sub-region shows at least three velocity components, indicating interaction among multiple substructures. Its field is much less than the average Zeeman field in molecular clumps, suggesting that the DCF value of the field there may be an underestimate. Numerical simulation of filamentary cloud formation shows that filamentary substructures can strongly perturb the magnetic field. DCF and true field values in the simulation are compared. Pre-stellar cores are observed in B211 and are seen in our simulation. The appendices give a derivation of the standard DCF method that allows for a dispersion in polarization angles that is not small, present an alternate derivation of the structure function version of the DCF method, and treat fragmentation of filaments.

\end{abstract}


\section{Introduction}

Filamentary structures have been found at almost all size scales in the Galaxy. Massive, long ﬁlamentary dark clouds are commonly found inside giant molecular clouds \citep[GMCs; e.g.][and references therein]{ber07,and14}, such as the dark clouds L1495 in the Taurus cloud complex \citep[e.g.][]{cha11} and the Serpens South cloud in the Serpens region \citep[e.g.][]{dha18}. Filamentary clouds of 4 to 6 pc length are common, and possibly longer than 10 pc. Some of these clouds are dark at infrared wavelengths. The line-width size relation observed for molecular gas indicates that the thermal Mach number would exceed 10 at such size scales. The long-term survival of these filamentary structures requires a reinforcing mechanism. As shown in the ideal magnetohydrodynamical (MHD) simulations of \citet{li19}, a moderately strong, large-scale magnetic field (\alfven Mach number, $\ma\sim 1$) can provide such a mechanism. In the weak-field model with $\ma =10$, the appearance of molecular clouds is clumpy, rather than the long and slender filamentary clouds seen in moderately strong field models. High resolution images of massive molecular clouds from the \textit{Herschel space telescope} reveal complex ﬁlamentary substructures \citep[e.g.][]{and14}. The characteristic inner width of molecular filaments found with \textit{Herschel} is about $\sim 0.1$ pc \citep{arz11, arz19}. Dense cores, where stars form, are located along or at the intersections of some of these ﬁne substructures \citep[e.g.][]{kon15,taf15}. From these observations of molecular cloud structures at diﬀerent size scales, one can visualize an evolutionary sequence of star formation starting from highly supersonic, magnetized GMCs, continuing on to ﬁlamentary dark clouds that form within them, and then on to ﬁner ﬁlamentary substructures. Fragmentation of these ﬁlamentary structures and substructures leads to the clumps and dense cores that form protostellar clusters and protostars. Knowing the physical conditions inside filamentary clouds would provide crucial information on the formation of filamentary substructures and dense cores, and on the origin of the initial mass function (IMF) and the star formation rate. Particularly important is the characterization of the physical properties of transcritical filamentary structures whose mass per unit length is within a factor of $\sim$2 of the critical line mass ${M_{\rm crit,\,th,\,\ell}}=2\, c_{\rm s}^2/G $ of nearly isothermal cylindrical filaments \citep[e.g.][]{ostr64,inu97}, where  $c_{\rm s}$ is the isothermal sound speed. Indeed, \textit{Herschel} observations suggest that transcritical filamentary structures dominate the mass function of star-forming filaments and that their fragmentation may set the peak of the prestellar core mass function and perhaps ultimately the peak of the IMF \citep[][]{and19}. In this paper, we report the results of polarimetric observations of the pristine section B211 of one such transcritical filament, the Taurus B211/B213 filament, using the High-resolution Airborne Wideband Campera plus (HAWC+) onboard Stratospheric Observatory For Infrared Astronomy (SOFIA). We determine the magnetic field structure inside a filamentary cloud with filamentary substructures.

The filamentary cloud L1495 is located in the Taurus molecular cloud at a distance of about 140 pc \citep{eli78}. Using their H-band polarization observation and the Davis-Chandrasekhar-Fermi (DCF) method \citep{dav51,cha53}, \citet{cha11} estimated the plane-of-sky (POS) magnetic field strength to be $10 - 17$ $\mu$G in the low density regions near the L1495 cloud, including the B211 region, and $25 - 28$ $\mu$G inside the cloud. From observations of $^{12}$CO and $^{13}$CO, they find that the velocity dispersion is $0.85 - 1.16$ km s$^{-1}$. Their observations have a resolution of 0.135 pc. The mean surface density is about $N({\rm H_2}) \sim 1.45\times10^{22}$~cm\ee\ \citep{pal13}.  Using the density estimated by \citet{hac13} and the measured velocity dispersion $\sim 1$ km~s\e\ cited above, the \alfven Mach number of the long filamentary cloud is about 2.7. Combining the polarization observations of \citet{hei00}, \citet{hey08}, and \citet{cha11}, \citet{pal13} found that the large-scale mean field direction is almost orthogonal to the cloud axis in B211/B213 and roughly parallel to faint striations seen in both CO and \textit{Herschel} data. There is also some kinematic evidence that the B211/B213 filament is embedded in a sheet- or shell-like ambient cloud and in the process of accreting mass from this ambient cloud \citep{shi19}, perhaps through the magnetically-aligned striations. Is it possible that the magnetic field pierces straight through the cloud? If so, then the picture of the formation of filamentary clouds is simple: gas is simply gathered into the cloud along approximately straight field lines.

A portion of the Taurus/B213 filament was recently mapped with JCMT-POL2 as part of the BISTRO project \citep{esw21}, but the corresponding $850\, \mu$m polarization data only constrained the magnetic field toward the dense cores within the filament. Other recent high-resolution polarization observations of magnetic field structures within more massive molecular clouds, such as Vela C \citep{sol13,sol17,dal19} and M17 SWex \citep{sug19}, show that magnetic fields inside dense filamentary clouds with complex substructures are not simple. Magnetic fields inside clouds can have large deviations from the large-scale field orientation. Within the $\sim 0.8$ pc outer diameter measured on \textit{Herschel} data, the B211/B213 filament system has a characteristic half-power width of $\sim 0.1$ pc and exhibits complex filamentary substructures \citep{pal13,hac13}. \citet{hac13} found that the B211 region has a mass of 138 M$_\odot$ and is roughly 2 pc long. From their  C$^{18}$O intensity map, the width of the B211 region encompassed by the lowest C$^{18}$O contour is about 0.3 pc. \citet{hac13} identified multiple velocity components in C$^{18}$O at different locations in B211-B213 with separations as large as about 2 km s$^{-1}$. \citet{taf15} find that the relative velocities between filamentary substructures in the filamentary cloud range over 2.2 km s$^{-1}$, possibly implying that the substructures are converging at high velocity. B211 is very bright in both C$^{18}$O and SO and has intense dust millimeter emission.  The gas in B211 has an unusually young chemical composition and lack of young stellar objects, indicating that this region is at a very early state of evolution \citep{hac13}. This region is therefore particularly suitable for the study of magnetic field structures in filamentary clouds without any confusion from protostellar activity.

In the high-resolution infrared dark cloud (IRDC) simulation by \citet{li19} using the adaptive mesh refinement code ORION2, a long filamentary cloud is formed in a moderately strong magnetic field environment, even though the thermal Mach number was 10. The long filamentary cloud created in the simulation has filamentary substructures similar to those in L1495. The simulation may therefore provide unique information on the physical environment inside filamentary clouds and on how they form. In the simulation, the large scale magnetic field is approximately perpendicular to the cloud axis, similar to the case in L1495. However, the small-scale magnetic field inside the simulated cloud, which has a width similar to that of B211, has a chaotic structure. Until now, there has never been a polarimetric observation with a resolution and a sensitivity high enough to peer into a filamentary cloud with no star formation. This motivates us to map a portion of L1495 to determine the field morphology inside the cloud and thereby gain an understanding of the physical environment inside such a cloud.

We report in this paper our observations of the filamentary cloud L1495/B211 using the recently optimized HAWC+ polarimeter on SOFIA to probe the complex magnetic field inside a slender filamentary cloud with complex filamentary substructures. The observation using HAWC+ polarimeter, the data reduction method, and the results are presented in Section 2. In Section 3, we investigate the physical state of the B211 region from observation. Using the DCF method, we estimate the magnetic field strength in Section 3.1 with the aid of recent C$^{18}$O (1-0) line emission data from the IRAM 30m telescope. In Section 3.2, we study the relation between the inferred magnetic field from HAWC+ observation and the surface density contours. In Section 4, results of our numerical simulation of filamentary clouds is used to provide insights into the physical state of B211. Our conclusions are presented in Section 5.

\section{Observations}
\subsection{SOFIA HAWC+ mapping observations and data reduction methods}
\label{sec:hawc}
L1495 was observed (ID: 07\_0017, PI: Li, P.S.) at 214 \um~($\Delta \lambda = 44$ \um, full width at half maximum, FWHM) using HAWC+ \citep{vai07,dow10,har18} on the $2.7$-m SOFIA telescope. HAWC+ polarimetric observations simultaneously measure two orthogonal components of linear polarization arranged in two arrays of $32 \times 40$ pixels each, with a detector pixel scale of $9$\farcs$37$ pixel$^{-1}$ and beam size (FWHM) of $18$\farcs$2$ at 214 \um. At 214 \um, HAWC+ suffers of vignetting where five columns cannot be used for scientific analysis \citep{har18}, therefore the field-of-view (FOV) of the polarimetric mode is $4.2 \times 6.2$ sqarcmin. We performed observations using the on-the-fly-map (OTFMAP) polarimetric mode. This technique is an experimental observing mode performed during SOFIA Cycle 7 observations as part of the shared-risk time to optimize the polarimetric observations of HAWC+. Although we will focus on the scientific results of L1495, we here describe the high-level observational steps used in these these observations, where Sections \ref{subsec:OTFMAP_limitations} and \ref{subsec:OTFMAPZeroLevel} describe the details of the OTFMAP polarimetric mode.

We performed OTFMAP polarimetric observations in a sequence of four Lissajous scans, where each scan has a different halfwave plate (HWP) position angle in the following sequence: $5^{\circ}$, $50^{\circ}$, $27.5^{\circ}$, and $72.5^{\circ}$. This sequence is called `set' hereafter (Table~\ref{tab:table1}-column 9). In this new HAWC+ observing mode, the telescope is driven to follow a parametric curve at a nonrepeating period whose shape is characterized by the relative phases and frequency of the motion. Each scan is characterized by the scan angle, scan amplitude, scan rate, scan phase, and scan duration. The scan angle is the relative angle of the cross-elevation direction of the FOV of the scan with respect to north, where $0^{\circ}$ is North and positive increase is in the east of north direction (Table~\ref{tab:table1}-column 7). An example of the OTFMAP for total intensity observations of NGC 1068 using HAWC+/SOFIA is shown by \citet[][fig. 1]{lop18}. The OTFMAP polarimetic mode using HAWC+/SOFIA at 89 \um~has recently been successfully applied to the galaxy Centaurus A \citep{lop21}. A summary of the observations at $214$ \um~are shown in Table \ref{tab:table1}. We performed square scans (Table \ref{tab:table1}-column 8) at three different positions as shown in Table \ref{tab:table1} (column 5 and 6). After combining all scans, the full FOV is $20\times20$ sqarcmin. Although Table \ref{tab:table1} lists all performed observations for this program with a total executed time of 6.37h, only a final executed time of 4.73h (with a total on-source time of 4.40h) was used for scientific analysis. The removed sets listed in Table \ref{tab:table1} were not used due to loss of tracking during observations.

\begin{table*}
\caption{Summary of OTFMAP polarimetric observations}
\label{tab:table1}
\begin{tabular}{cp{1.4 cm}cccp{0.6 cm}p{1.6 cm}cc}
\hline
Date & Flight Number & Altitude & RA & DEC & Scan Time & Scan Angle & Scan Amplitude & \# Sets  \\ 
(YYYYMMDD) &  & (ft) & (h) & ($^{\circ}$) & (s) & ($^{\circ}$) &  (EL $\times$ XEL; \arcsec)  & used (removed) \\
\hline
20190904	&  F605  	&   41000	&	4.3036 & 27.5415 & 120	&   -30.0, 0.0, 23.7	&   300$\times$300  & 1 (2) \\
		&		&		&			&		& 60	&   -23.7			&   300$\times$300 & 1 \\
20190905 &  F606    &   42000, 43000 & 4.3036 & 27.5415 & 60 & -30.0, -26.8, 30, -20.5, -17.4, -14.2, -7.9, -4.7 & 300$\times$300 & 8 (1)  \\
20190907 &  F607   & 42000, 43000 & 4.3036 & 27.5415 & 120 & -30.0, -26.7, -23.7, -20.5, -17.4, -14.2, -11.0 & 300$\times$300 & 7 \\
20190910 & F608   & 42000, 43000  & 4.3075 & 27.4836 & 120 & -30.0, -26.8, -23.7, -20.5, -17.4, -14.2, -11.0, -7.8, -4.7, -1.6  & 300$\times$300 & 10 \\
20190918  & F611 & 42000, 43000   & 4.3036 & 27.5415 & 120 & -30.0, -28.8, -23.7, -20.5, -17.4, -14.2, -11.0, -7.9 &  300$\times$300 & 0 (8) \\
20191010 & F621 & 43000 & 4.3075 & 27.4836 & 120 & -30.0, -26.8, -23.7, 0.0, -7.1, -14.2, 30.0, 12.6, -4.7, 60.0 & 330$\times$330 & 10 \\
\hline
\end{tabular}
\end{table*}

We reduced the data using the Comprehensive Reduction Utility for SHARP II v.2.42-1 \citep[\textsc{crush};][]{kov06,kov08} and the \textsc{HAWC\_DRP\_v2.3.2} pipeline developed by the data reduction pipeline group at the SOFIA Science Center. Each scan was reduced by \textsc{crush}, which estimates and removes the correlated atmospheric and instrumental signals, solves for the relative detector gains, and determines the noise weighting of the time streams in an iterated pipeline scheme. Each reduced scan produces two images associated with each array. Both images are orthogonal components of linear polarization at a given HWP position angle. We estimated the Stokes $I,\,Q,\,U$ parameters using the double difference method in the same manner as the standard chop-nod observations carried by HAWC+ described in Section 3.2 by \citet{har18}. The degree ($P$) and position angle of polarization were corrected by instrumental polarization (IP) estimated using OTFMAP polarization observations of planets. We estimated an IP of $Q/I = -1.0$\% and $U/I = -1.4$\% at $214$ \um~respectively, with an estimated uncertainty of $\sim0.8\%$. The IP using OTFMAP observations are in agreement with the estimated IP using chop-nod observations. To ensure the correction of the position angle of polarization of the instrument with respect to the sky, we took each set with a fixed line-of-sight (LOS) of the telescope. For each set, we rotated the Stokes $Q$ and $U$ from the instrument to the sky coordinates. The polarization fraction was debiased \citep{war74} and corrected by a polarization efficiency of $97.8$\% at 214 \um. The final Stokes $I$ and its associated errors were calculated and downsampled to the beam size ($18\farcs20$). The final Stokes $Q,\, U,\, P$, position angle, polarized intensity ($PI$), and their associated errors were calculated and re-sampled to a super-pixel of $3\times3$ detector pixel size, which corresponds to a re-sampled pixel size of $28\farcs1$
(or 0.019 pc at the distance of the cloud). This super-pixel was chosen to optimize the signal-to-noise ratio (SNR), obtain statistically independent measurements and significant polarization measurements without compromising spatial resolution for the data analysis. 

\subsection{HAWC+ OTFMAP polarization: advantages and limitations}
\label{subsec:OTFMAP_limitations}

Several advantages and limitations are found with the OTFMAP polarization mode. The advantages are the reduction of overheads and radiative offsets when compared with the chop-nod technique. The overheads of the OTFMAP are estimated to be 1.1 in comparison with the typical overheads of 2.7 by the chop-nod technique, which shows an improvement by a factor $\ge2$. This improvement is due to OTFMAP constantly integrating with the source on the FOV while covering off-source regions to estimate the background levels, and observing overheads. For the OTFMAP method, the telescope is always on-axis, without chopping the secondary mirror as it is in the chop-nod technique. Thus, the radiative offset is not present and the sensitivity of the observations was estimated to improve by a factor of 1.6. The OTFMAP technique provides a larger map area when compared to the chop-nod technique. Our observations were taken at three different positions covering a FOV of $10\times10$ and $11\times 11$ sqarcmin, yield a final FOV of $20\times20$ sqarcmin. Note the advantage of the large FOV by the OTFMAP when compared with the single $4.2\times6.2$ sqarcmin by the chop-nod technique. 

The limitation of the OTFMAP technique lies in the recovering of large-scale diffuse and faint emission from the astrophysical objects. This is a result of the finite size of the array, variable atmosphere conditions, variable detector temperatures, and the applied filters in the reduction steps to recover extended emission. We applied several filters using \textsc{crush} to recover large-scale emission structures of L1495 while paying close attention to any change that may compromise the intrinsic polarization pattern of the astrophysical object. We conclude that the \textsc{faint} filter with a number of 30 iterations from \textsc{crush} can recover large-scale emission structures larger than the Band E FOV from our observations of L1495. The \textsc{faint} option applies filtering to the timestreams and extended structures to recover fluxes with SNR $<10$ in a single scan. In addition, the number of rounds are such as that the iterative pipeline is able to recover large-scale structures without introducing additional artificial structures not identified in the \textit{Herschel} images. In general, the noise increases as a function of the length, $L$, of the extended emission as $\sim L^{2}$.  We force each individual scan produced by \textsc{crush} to have a pixel scale of $3\times3$ detector pixels ($28\farcs$1), which increase the SNR of each scan by a factor of 3 helping to recover larger and fainter structures.

\subsection{HAWC+ OTFMAP polarization: zero-level background}
\label{subsec:OTFMAPZeroLevel}

An important step is the estimation of the zero-level background of the observations. We remind the reader that HAWC+ measures the power of the emissive and variable atmosphere and the astrophysical object. The data reduction scheme described above produces regions of negative fluxes in areas of extended and low surface brightness due to the similar levels of noise and astrophysical signal. Thus, it is of great importance to characterize and estimate the zero-level background across the  observations of L1495, because there is a potential loss of flux that requires to be estimated and added to the full image.

\begin{figure*}
\includegraphics[angle=0,scale=0.7]{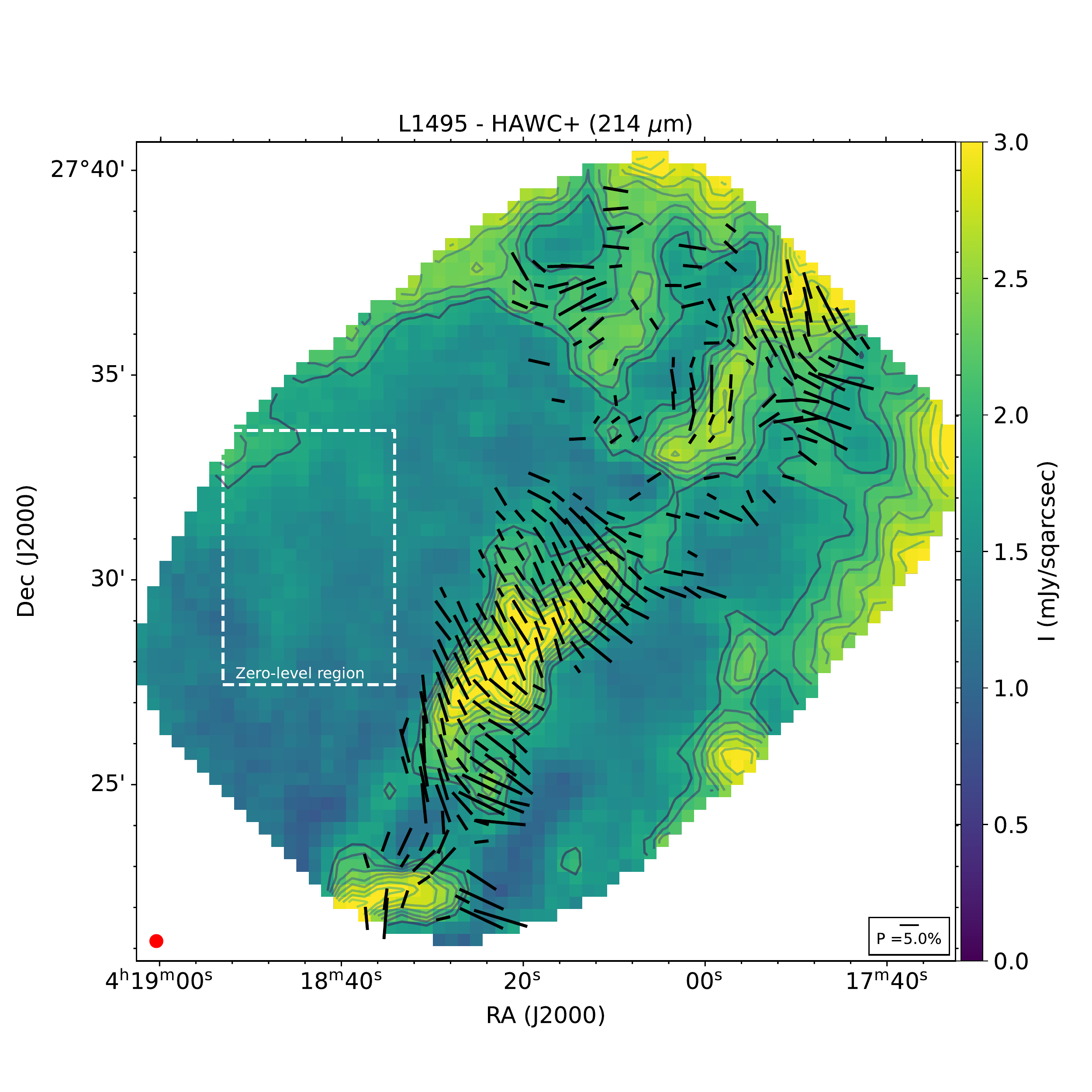}
\caption{Total surface brightness at 214 $\mu$m of L1495 within a $1200 \times 1200$ sqarcsec region using the OTFMAP observations. Contours start at $45\sigma$ and increase in steps of $5\sigma$, with $\sigma = 0.032$ mJy/sqarcsec. Although quality cuts have been performed to cut edge effects, there are still structural artifacts at the corners of the map (specially in the West region) due to the low coverage at the edges of the map. Polarization measurements (black lines) have been rotated by $90^{\circ}$ to show the inferred magnetic field morphology. The length of the polarization vector is proportional to the degree of polarization. Only vectors with $P/\sigma_{P} \ge 2$ are shown. A legend with a $5\%$ polarization and the beam size ($18\farcs2$) are shown. The zero-level background region (white dashed line) described in Section \ref{subsec:OTFMAPZeroLevel} is shown.
\label{fig:polmap}}
\end{figure*}

We have determined and corrected the zero-level background of our observations as follows. Using \textit{Herschel} images at 160 and 250 \um~from the \textit{Herschel} Archive\footnote{\textit{Herschel} archive: \url{http://archives.esac.esa.int/hsa/whsa/}}, we identified a region in the sky where the fluxes of an individual pixel of size 28\farcs1 are below the sensitivity of HAWC+ at 214 \um. This area is shown in Fig. \ref{fig:polmap}, which is located in a common region for all scans across the multiple flights. The size of this region is chosen to be equal to the HAWC+ FOV at Band E, i.e. $4.2\times6.2$ sqarcmin. The size of the background region was chosen to be the same as if the observations were performed using the chop-nod observing mode. The size of this region does not influence the estimation of the zero-background level, rather the location and the surface brightness do. Then, for both arrays and each HWP position angle produced from the first step by \textsc{crush}, we estimate the mean and standard deviation within the zero-level region. To remove negative values across the image, the mean is added to all pixels in each scan and HWP position angle. After this step, the same reduction procedures as described in Section \ref{sec:hawc} are followed. Using archival chop-nod and OTFMAP observations of well-known objects, e.g. 30 Doradus and OMC-1, and applied the same methodology, we reached similar conclusions and methodologies, while the polarization pattern was shown to be consistent between reduction schemes. Finally, we computed the SED of the source using  $70-500$ \um~\textit{Herschel} images and estimated the expected flux at $214$ \um. We estimated that the fluxes from the zero-level background corrected image are within $8$\% from the expected flux from the \textit{Herschel} SED, which is within the flux calibration uncertainty of HAWC+ of $\le15$\% provided by the SOFIA Science Center.

\begin{figure*}
\includegraphics[angle=0,scale=0.7]{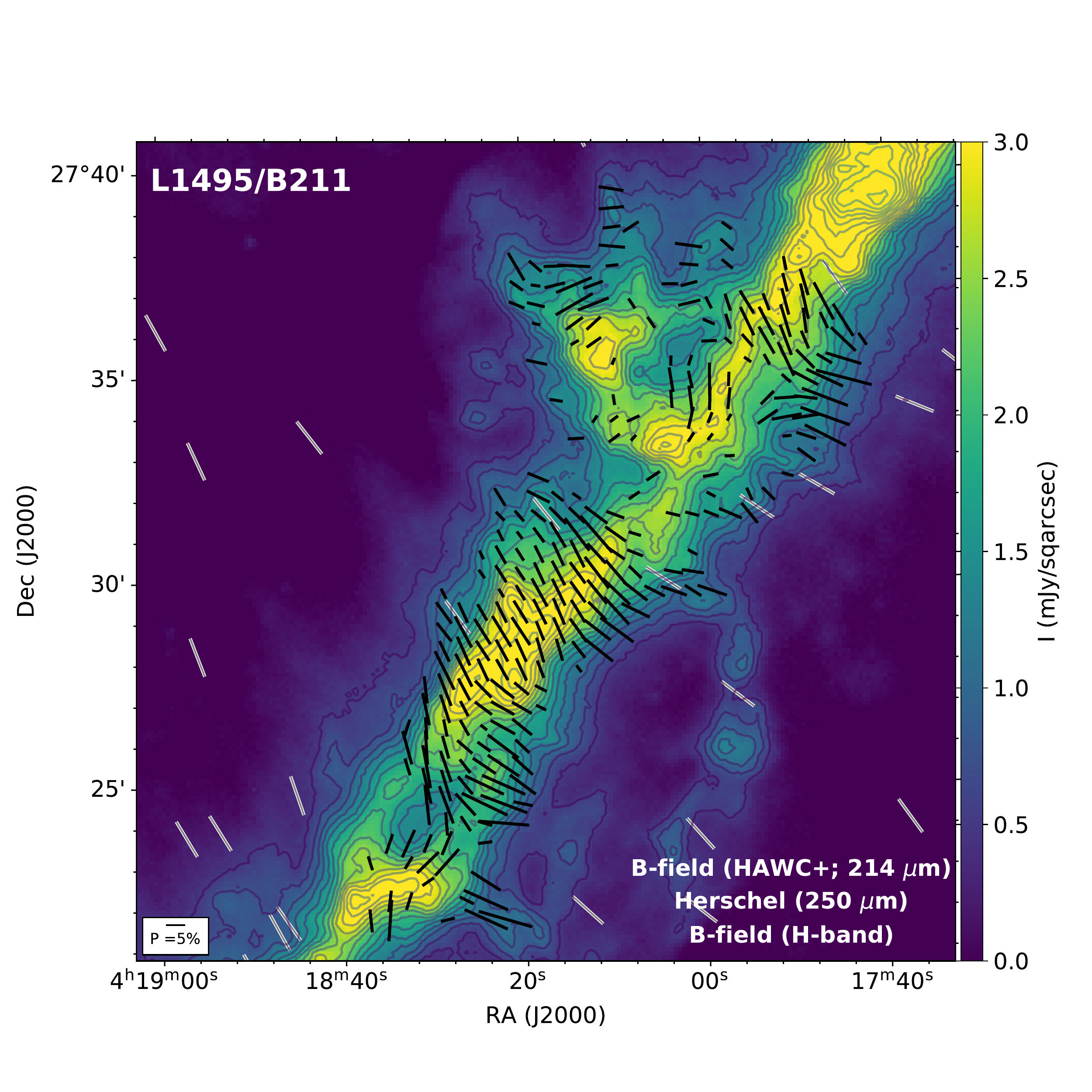}
\caption{The 250 $\mu$m total surface brightness from \textit{Herschel}/SPIRE (color scale) with the magnetic field morphology as inferred from the 214 $\mu$m HAWC+ (black lines) and H-band \citep[grey lines;][]{cha11} observations. Contours start at $0.5$ mJy/sqarcsec and increase in steps of $0.25$ mJy/sqarcsec.
\label{fig:polmapcomb}}
\end{figure*}
Although the zero-level background region has low surface brightness, the polarization may be high and contaminate the astrophysical signal after the zero-level background correction. Here, we estimate the contribution of the zero-level background to the polarization measurements. As mentioned above, the mean and standard deviation within the zero-level background region was estimated for each array and HWP position angle. Using the double difference method \citep[Section 3.2 by][]{har18}, we estimate the Stokes $Q$ and $U$ and their uncertainties by spatial averaging within full FOV of the zero-level background region. Then, the Stokes $Q$ and $U$ were corrected by instrumental polarization, and $P$ and position angle were estimated and corrected by bias and polarization efficiency. Finally, the $P$ and position angle of the zero-level background region were estimated to be $8.5\pm3.5$\% and $42\pm8^{\circ}$, respectively. The minimum detectable flux from Stokes $I$ is estimated to be $3\sigma_{I} = 0.096$ mJy/sqarcsec, which corresponds to a polarized flux of $3\sigma_{PI} = 0.008$ mJy/sqarcsec using $P = 8.5$\% . From our polarization measurements with $P/\sigma_{P} \ge 2$, we estimate a median polarized flux of $0.055$ mJy/sqarcsec. Thus, the zero-level background correction contributes a median of $\sim14$\% to the polarized flux in our measurements.

\subsection{HAWC+ polarization map and orientation of magnetic fields}
\label{sec:polarization}
Fig. \ref{fig:polmap} shows polarization measurements projected onto the total surface brightness at 214 $\mu$m of the $1200\times1200$ sqarcsec region of L1495/B211 that we observed. The polarization measurements have been rotated by $90^{\circ}$ to show the inferred magnetic field morphology. {\it All polarization angles (PAs) cited in this paper have been rotated in this manner.} Only polarization measurements with $P/\sigma_{P} \ge 2$ are shown \citep{war74}. The length of the polarization lines are proportional to the degree of polarization, where a $5$\% polarization measurement is shown as reference. Images had edges artifacts due to the sharp changes in fluxes  and limited number of pixels. The final 214 \um~HAWC+ polarization measurements contain pixels with the following quality cuts:  1) pixels with a scan coverage $\ge30$\% of all observations, 2) pixels which Stokes I measurements have an uncertainty $\ge2.5\sigma_{I}$, where $\sigma_{I}$ is the minimum uncertainty in Stokes I, 3) pixels with a surface brightness of $\ge1$ mJy/sqarcsec, 4) pixels with $P\le30$\% given by the maximum polarized emission found by \textit{Planck} observations \citep{pla13}, and 5) polarization measurements with $P/\sigma_{P} \ge 2$. We find that 14\% (40 out of 282) of the measurements are within $2 \le P/\sigma_{P} \le 3$.

\begin{figure}
\includegraphics[scale=0.47]{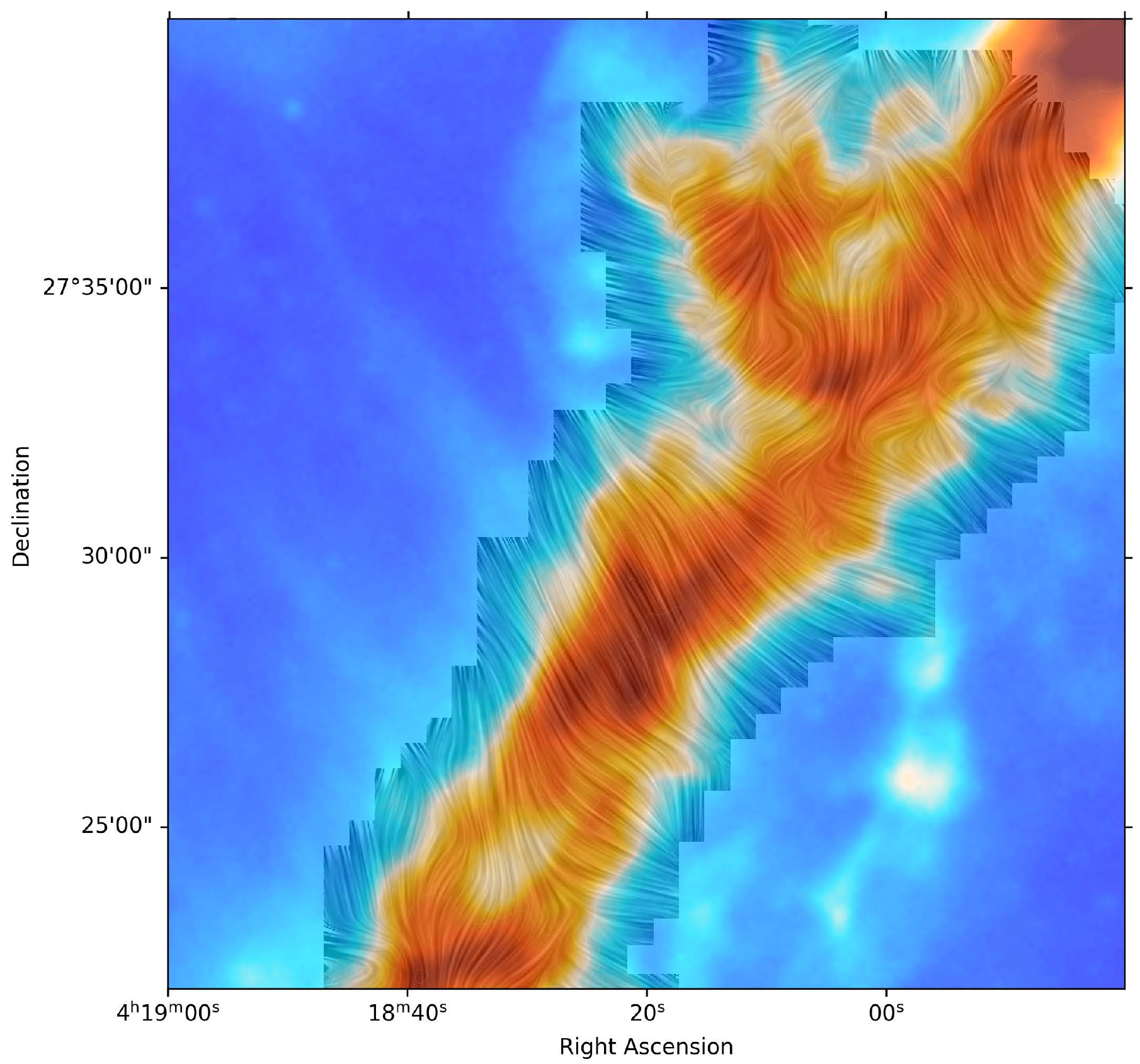}
\caption{The inferred magnetic field orientation from HAWC+ polarization observations at 214 \um~of the B211 filament is shown using the linear integral convolution algorithm \citep[LIC;][]{cab93}.   Same polarization measurements as Figure \ref{fig:polmap}, a resample scale of 20, and a contrast of 4 were used. The colorscale shows the $250 \mu$m total intensity image from the \textit{Herschel} Gould Belt survey as shown in Figure \ref{fig:polmapcomb}.
\label{fig:lic}}
\end{figure}

\begin{figure}
\includegraphics[scale=0.75]{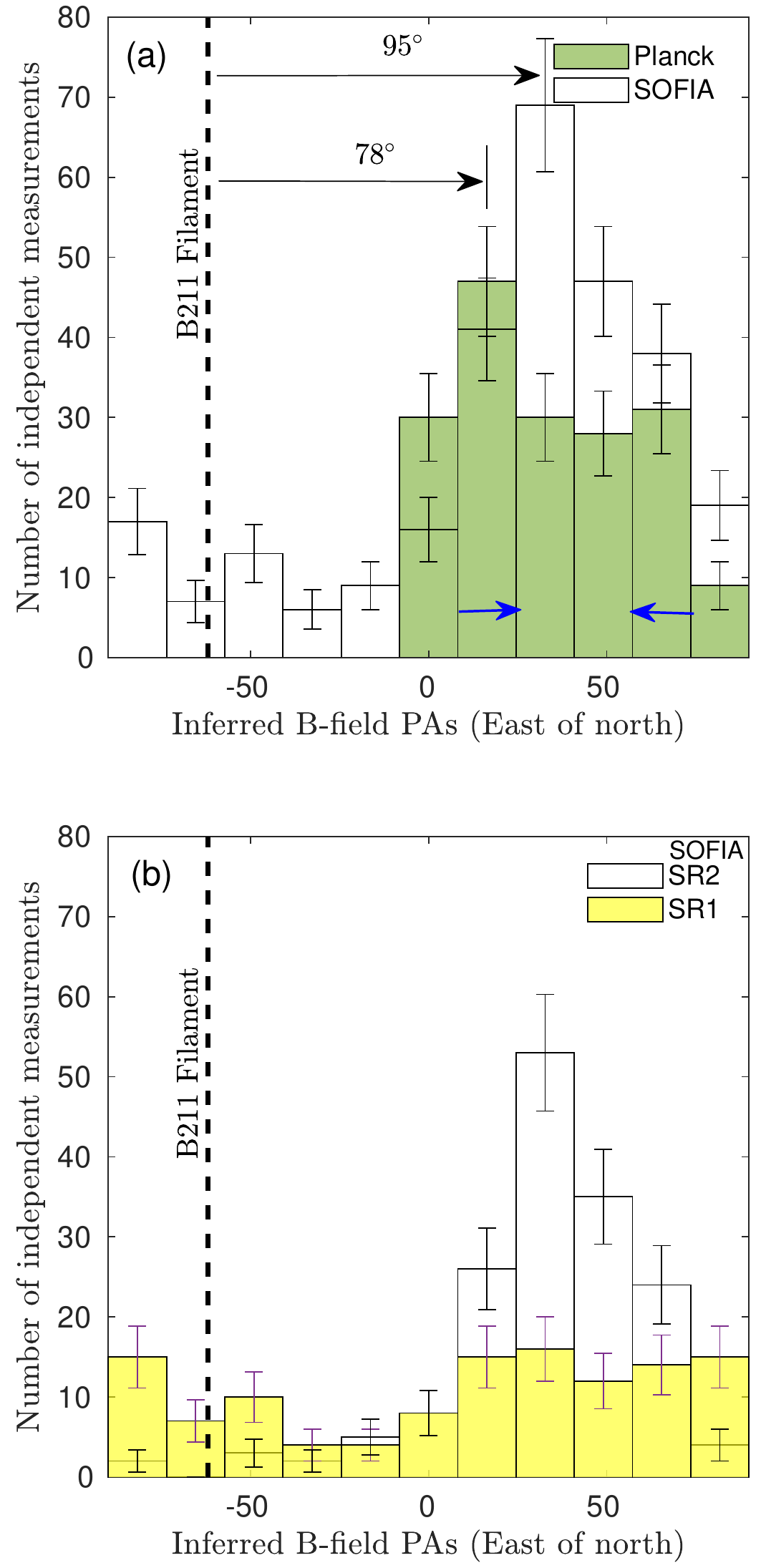}
\caption{(a) The histograms of all the inferred magnetic field orientations from HAWC+ polarization observations of the B211 filament at resolution of 28.1 arcsec shown in Fig. \ref{fig:polmapcomb} (empty) and from Planck polarization observations (green) of the entire Taurus/B211 region at resolution of 10 arcmin shown in Fig. \ref{fig:all_data}. The six Planck polarization measurements within the FOV of our observations are inside the two bins $24^\circ - 57^\circ$ marked by the two blue arrows. The bin width of 16.4$^\circ$ is chosen to be larger than the measurement uncertainty of the PA. The vertical dashed line indicates the orientation of the B211 filament at PA = $-62^\circ$ equivalent to $+118^\circ$  (\citet{pal13}). The angle differences of the peaks of the two sets of histograms from the PA of the B211 filament are shown. (b) The histograms of inferred magnetic field orientations from HAWC+ polarization observations of sub-regions SR1 (yellow) and SR2 (empty).
\label{fig:observed_pa}}
\end{figure}

\begin{figure*}
\includegraphics[scale=0.55]{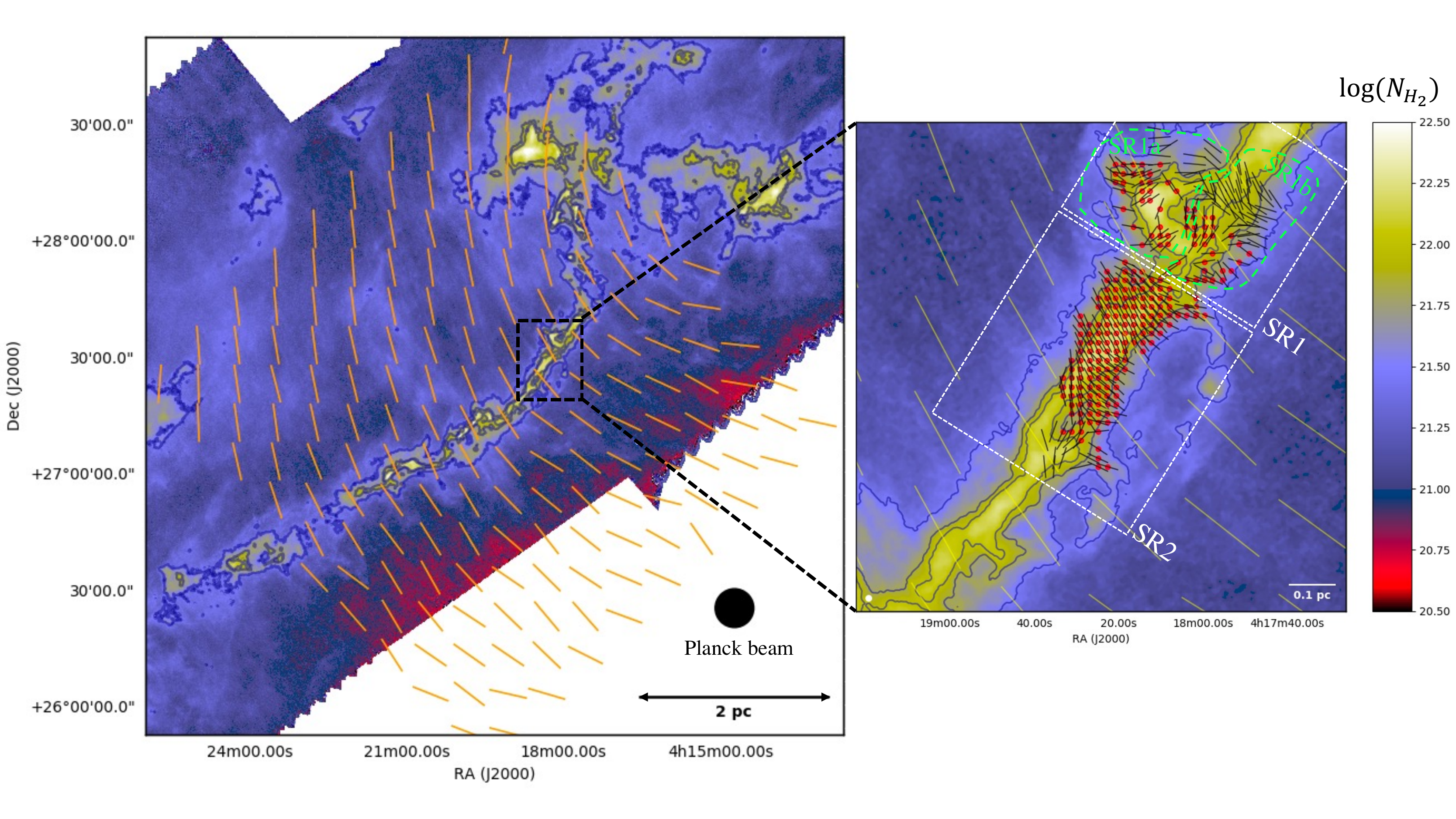}
\vspace{-0.35cm}
\caption{ {\it Left:}  Column density map from {\it Herschel} Gould Belt survey data at 18.2$\arcsec$ resolution \citep{pal13,and10}, with magnetic field vectors derived from {\it Planck} polarization data at 10$\arcmin $ resolution  \citep{pla13} displayed in orange/yellow (and spaced every 10$\arcmin $). Contours are $N({\rm H_2}) = 3 \times 10^{21}$, $6.7 \times 10^{21}$, and $10^{22}$~cm$^{-2}$. {\it Right:} Blow-up of the left image in the area mapped with SOFIA/HAWC$+$. Yellow vectors are from {\it Planck} and are here spaced by half a beam (5$\arcmin $). Smaller black segments show the magnetic field vectors derived from HAWC$+$ at 28$\arcsec $ resolution. The solid red circles mark positions where both significant HAWC$+$ polarization measurements and C$^{18}$O line data from the IRAM 30m telescope are available. The two sub-regions (SR1 and SR2) for which a DCF analysis has been carried out are marked by white dotted rectangles; the two components of SR1 (SR1a and SR1b) are outlined by green dashed contours.
\label{fig:all_data}}
\end{figure*}

In Fig. \ref{fig:polmapcomb} we over-plotted the magnetic field orientations from near-infrared H-band observations obtained by \citet{cha11}. Note that the inferred magnetic field from the H-band arises from dichroic absorption, while our 214 $\mu$m measurements arise from dichroic emission. We detect many multiple structures of magnetic field over just 0.82 pc region inside the 2-pc-long B211 region from the HAWC+ polarization mapping at smaller scales $28\farcs1$ compared to the lower resolution observation from the H-band and Planck observations (see Fig. \ref{fig:all_data}).  We note that the magnetic field of the lower half of the observed area is more uniform and close to the perpendicular direction of the filamentary cloud. From the \textit{Herschel} intensity map, this part of B211 appears to have two filamentary substructures crossing each other in an x-shape appearance near RA of $4^{\rm h}18^{\rm m}20^{\rm s}$ and DEC of $27^{\circ}29'$. The two structures may be spatially nearby and appeared to be overlapped along the LOS. The magnetic field could be a combined result of the overlapping projection. In the upper half, the magnetic field structure appears as a highly non-uniform chaotic state. It is consistent with the turbulent appearance of the underlining intensity map that there could be three tangling filamentary substructures at this location as identified by \citet{hac13}. The deviation of polarization angle is large from the uniform large scale field direction indicated by the near-infrared H-band and Planck observations. Fig. \ref{fig:lic} is a line integral convolution \citep{cab93} plot of the inferred magnetic field from the HAWC+ polarization observations.

The histogram of the $B$-field PA distribution of all the polarization measurements detected with $P/\sigma_{P} > 2$ is shown in Fig. \ref{fig:observed_pa}a. The peak is near $30^\circ$, which is very close to the large scale mean magnetic field orientation of $26^\circ \pm 18^\circ$ and is nearly orthogonal to the L1495 filamentary cloud axis at $118^\circ \pm 20^\circ$ \citep{pal13}. If the PAs have a range approaching 180\degree, then the dispersion can depend on the choice of $\theta=0^\circ$ since PAs near +90\degree\ can be flipped to -90\degree\ by a change in the orientation of 0\degree. In this paper we evaluate the dispersion in PAs by choosing 0\degree\ to be consistent with the minimum dispersion in PAs. We find that the PAs in Fig. \ref{fig:observed_pa} have a dispersion of $36.1^\circ$, indicating that the small-scale magnetic field is strongly perturbed inside the cloud compared with the large scale field. Basically, the inferred B-field are pointing at all directions in the northwestern side region of 0.82 pc in size.  The large-scale field PA distribution from Planck is also plotted in Fig. \ref{fig:observed_pa}a for direct comparison. Note that most of the magnetic field orientations from Planck are located far from the B211 region (see Fig. \ref{fig:all_data}) and have a dispersion of $37^\circ$. The several Planck polarization orientations inside the observed B211 region (indicated by a black dash-line box in Fig. \ref{fig:all_data}) are all inside the two bins between $24^\circ - 57^\circ$ at the peak of the Planck distribution and close to the peak of the mean magnetic field inside B211 from HAWC+. The resolution of the large-scale field from Planck is $\sim 0.4$ pc, which is about 21 times the size of the super-pixel that we adopted for the HAWC+ results. In Section \ref{sec:dcf_sim}, we use a numerical simulation to discuss how the resolution of a polarization map can affect the interpretation of magnetic field structures.

\subsection{IRAM 30m observations}
\label{sec:iram}

C$^{18}$O(1--0) mapping observations of a portion of the B211 field imaged with HAWC$+$ were carried out with the Eight MIxer Receiver (EMIR) receiver on the IRAM-30m telescope at Pico Veleta (Spain) in April 2016, as part of another project (Palmeirim et al. in preparation). At 109.8 GHz, the 30 m telescope has a beam size of $\sim$23$\arcsec$ (HPBW), a  forward efficiency of 94\%, and a main beam efficiency of 78\% \footnote{ \url{http://www.iram.es/IRAMES/mainWiki/Iram30mEfficiencies}}. As backend, we used the VESPA autocorrelator providing a frequency resolution of 20 kHz, which corresponds to a velocity resolution of $\sim$0.05 km s$^{-1}$ at 110 GHz. The standard chopper wheel method\footnote{Chopper wheel method used in IRAM-30m can be found at \url{https://safe.nrao.edu/wiki/pub/KPAF/KfpaPipelineReview/kramer_1997_cali_rep.pdf}} was used to convert the observed signal to the antenna temperature $T_{\rm A}^*$ in units of K, corrected for atmospheric attenuation. During the  observations, the system noise temperatures ranged from $\sim$85 K to $\sim$670 K. The telescope pointing was checked every hour and found to be better than $\sim$3$\arcsec$ throughout the run.

\begin{figure*}
\includegraphics[scale=0.55]{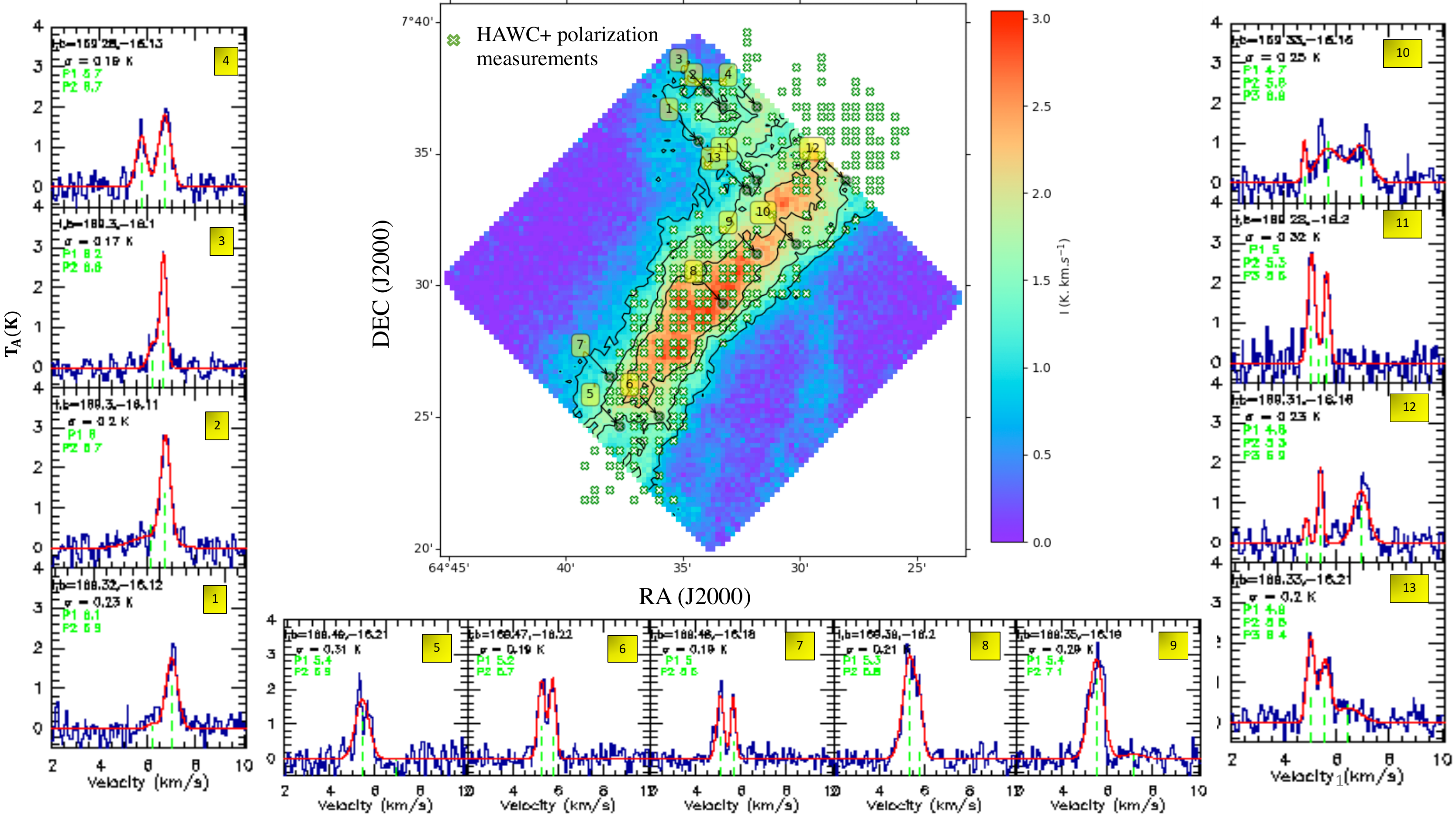}
\caption{C$^{18}$O(1--0) integrated intensity map over all channel velocities from 4.5 to 7 km s$^{-1} $. The contours correspond to 30\%, 50 \%,  and 70\% of the maximum integrated intensity (3 K km s$^{-1}$). Markers ($x$) in green indicate positions where statistically significant polarization measurements were obtained with HAWC+. Representative C$^{18}$O(1--0) spectra observed with the IRAM 30m telescope at selected positions in the field are shown to the left, bottom, and right of the map.
\label{fig:illu_vel}}
\end{figure*}

\section{Physical state of B211 region inferred from observations}
\label{sec:comparison}

\subsection{Magnetic field strength in B211}
\label{sec:dcf}

We derive magnetic field strengths using the DCF method based on the observed velocity dispersion, surface density (which provides an estimate of the gas density), and the dispersion of polarization angles. We used IRAM 30m C$^{18}$O data to derive the velocity dispersion and a HAWC+ polarization map to derive the dispersion in polarization angles. The density is based on {\it Herschel} column density data published in the literature. We describe the detailed methods below. The observed and derived parameters are summarized in Table \ref{tab:dcf_est}.

\subsubsection{Davis-Chandrasekhar-Fermi Method}
\label{sec:DCF}

The \citet{dav51}-\citet{cha53} method (hereafter the DCF method) allows one to infer the strength of the POS magnetic field from observations of the fluctuations in the polarization angle (PA) and is discussed extensively in Appendix A. The mean POS field is denoted by $B_0$ and is related to the mean 3D field by
\beq
B_0=B_{0,\rm 3D}\cos\gamma,
\label{eq:3d}
\eeq
where $\gamma$ is the angle between $\vecB_0$ and the plane of the sky. In the original DCF method, the mean POS magnetic field was determined over the entire field of view over which the PAs were measured. An expression for the value of this field that is valid for larger dispersions than the original DCF result and is related to a result obtained by \citet{fal08} is given in equation (\ref{eq:bo2.5}):
\beqa
B_0&=&\fdcf \,\frac{(4\pi\rho)^{1/2}\sigma_V}{\tan\sigma_\theta},\\
&=& 0.383 \sqrt{n(\htwom) }\, \frac {\sigma_V   }{\tan\sigma_{\theta}}  ~~~\mu {\rm G},
\label{eq:bpos2}
\eeqa
where $n(\htwom)$ is the number density of H$_2$ molecules in cm$^{-3}$, $\sigma_V$ is measured in km s\e, and $\sigma_{\theta}$ is the dispersion in the orientation of the magnetic field orientations, and where we have set the factor $\fdcf$, which corrects for the approximations made in deriving the DCF relation, to be 0.5 based on the results of \citet{ost01}. Comparison with numerical simulations confirms that this formula (with $\tan\sigma_\theta$ replaced by $\sigma_\theta$ in radians under the assumption that $\sigma_\theta$ is small) is valid when $\sigma_\theta \leq 25^\circ$ \citep{ost01}. The latter relation (with $\sigma_\theta$) is often used for larger dispersions, however.

A key step in the DCF method is to infer the dispersion in the field, $\sdb$, from the dispersion in PAs, $\sigma_\theta$. A complication is that the field angles (FAs) can range over $-180^\circ$ to $+180^\circ$, whereas the PAs are restricted to the range $-90^\circ$ to $+90^\circ$. As a result, the direction of the implied field depends on the choice of zero angle for the PAs. A field angle of 60\degree\ if 0\degree\ is vertical becomes -30\degree if the coordinate system is rotated 90\degree\ counterclockwise. Since the magnitude of the PA depends on the choice of coordinate system, it follows that the value of $\sigma_\theta$ does also. In this paper, we adopt the convention that we choose the coordinate system that minimizes the dispersion, $\sigma_\theta$, as recommended by \citet{pado01}. This becomes relevant only if some of the field angles differ by more than 180\degree, which generally occurs only if $\sigma_\theta$ is not small.

A second method of inferring the turbulent field strength from the spatial variation in the PAs is the structure function method introduced by \citet{hil09} and extended by \citet{hou09} and \citet{hou16}. This method is more general since it allows for a smooth variation in the orientation of the mean field. The structure function relates the PAs at different points
and is defined as
\beq
\langle\Delta\Phi(\ell)^2\rangle \equiv \frac{1}{N(\ell)} \sum\limits_{i=1}^{N(\ell)} [\Phi(\vecx) - \Phi(\vecx+\vecl)]^2,
\label{eq:sf0}
\eeq
where $\Phi(\vecx)$ is the PA at position $\vecx$, $\vecl$ is the displacement, and $N(\ell)$ is the number of polarization angle pairs with separation $\ell$. The structure function is related to the fluctuations in the magnetic field by equation (\ref{eq:omcos}) in the small angle approximation. In order to infer the field dispersion, $\Delta\Phi(\ell)$ is extrapolated to $\ell=0$, which gives $\Delta\Phi_0=\surd 2\sdb/B_\rms$. Unlike \citet{hil09}, we insert a factor $\fdcf$ into the result for $B_0$ and set $\fdcf=0.5$. Equation (\ref{eq:bo4}) then gives
\beq
B_0=0.383\,n(\htwom)^{1/2}\sigma_V\,\frac{(2-\Delta\Phi_0^2)^{1/2}}{\Delta\Phi_0}.
\label{eq:bpos3}
\eeq
The determination of $\sigma_\theta$ and $\Delta\Phi_0$ is discussed in section \ref{sec:ang_dispersion} below.

\subsubsection{IRAM 30m C$^{18}$O data and Velocity Dispersion}

Thanks to their high sensitivity, the IRAM 30m molecular line data highlight the kinematic complexity of the region mapped by HAWC$+$, with the presence of multiple velocity components. These multiple velocity components are consistent with the presence of filamentary substructures in this region as discussed by \citet{hac13}. The variety of observed C$^{18}$O(1--0) spectra is illustrated in Fig.~\ref{fig:illu_vel}, which shows clear changes in the number of velocity components and in overall centroid velocity as a function of position within and around the B211 filament. 

C$^{18}$O(1--0) molecular line data trace the kinematics of the gas and can be used to estimate the level of non-thermal motions due to turbulence in the region. As the C$^{18}$O(1--0) transition is usually optically thin, multiple peaks in the spectra, when present, likely trace the presence of independent velocity components as opposed to self-absorption effects. For better characterization of the different velocity peaks, we performed multiple Gaussian fits which allowed us to identify the centroid position of each velocity component where multiple components are observed. Comparing all of the C$^{18}$O(1--0) spectra observed in a given sub-region, it was possible to identify a dominant velocity component in each case. Table \ref{tab:dcf_est} provides the centroid velocity and velocity dispersion of the dominant and total velocity component in each sub-region for the significant HAWC$+$ detection where $P/\sigma_{P}\geq 2$, where $P$ represents the polarization degree (first row for each sub-region in Table~\ref{tab:dcf_est}). 

The centroid velocities of the relevant velocity components range 
from 5.4~km s$^{-1}$ to 5.9~km s$^{-1}$, and the associated line-of-sight 
velocity dispersion range from $\sim$0.2 to $\sim$0.3 km s$^{-1} $ for the dominant components and from $\sim$0.4 to $\sim$0.5 km s$^{-1} $ if all velocity components are considered.

\subsubsection{Dispersion in polarization angles from HAWC+}
\label{sec:ang_dispersion}
The SOFIA HAWC+ polarization data reveal a strongly perturbed structure of the magnetic field in the B211 region. In this work, we estimated the dispersion in polarization angles using independent polarization measurements in two sub-regions of B211, called SR1 and SR2 in Fig. \ref{fig:all_data}. The motivation for this division is that the region with polarization detections (see Fig.~\ref{fig:polmap}) is clearly not homogeneous: the southeastern part (SR2) corresponds to a segment of the B211 main filament, while the northwestern part (SR1) is an interaction region where material associated with the striations seen in the {\it Herschel} data meet the main filament  \citep[cf.][]{pal13}. Moreover, SR1 and SR2 correspond to different groups of C$^{18}$O velocity components, namely components \#9, \#12, \#11 for SR1, and components \#12, \#14 for SR2, in the analysis presented by \citet{hac13}. It is also apparent from Fig.~\ref{fig:polmap} that the dispersion of polarization angles is significantly higher in SR1 than in SR2. Taking into account measurement errors in our polarization data, we estimate  $\sigma_\theta$ as the weighted standard deviation of polarization angles in each sub-region: 
\begin{equation}
{\sigma_\theta^2  } = \frac{N}{N-1}\,\frac{1}{w}\, 
\sum\limits_{i=1}^N w_i \, (\theta_i - \bar{\theta}_w)^2 , 
\label{eq:sigmatheta}
\end{equation}
\noindent
where $N$ is number of independent measurements in each sub-region, $w_i = 1/\sigma_i^2 $ the weight of measurement $i$ given the measurement error in PA $\sigma_i$, $w = \sum\limits_{i=1}^N w_i$, and $\bar{\theta}_w = (1/w)\, \sum\limits_{i=1}^N w_i\, \theta_i$ is the weighted mean polarization angle in the sub-region. In sub-region~1 (SR1 from Fig. \ref{fig:all_data}), where there are 120 independent HAWC$+$ measurements, the dispersion in polarization angles is $54^{\circ}\pm 5^{\circ}$, which is a rather large value considering the main regime of applicability of the DCF method ($\sigma_\theta\leq 25^{\circ}$ -- see \citealp{ost01}). The error in the dispersion of polarization angles was estimated as  $\sigma_\theta / \sqrt{N}$. In sub-region 2 (SR2), there are 162 independent HAWC$+$ measurements and the dispersion in polarization angles is $20^{\circ}\pm 2^{\circ}$ (see Table \ref{tab:dcf_est}). 

\begin{figure}
\includegraphics[scale=0.45]{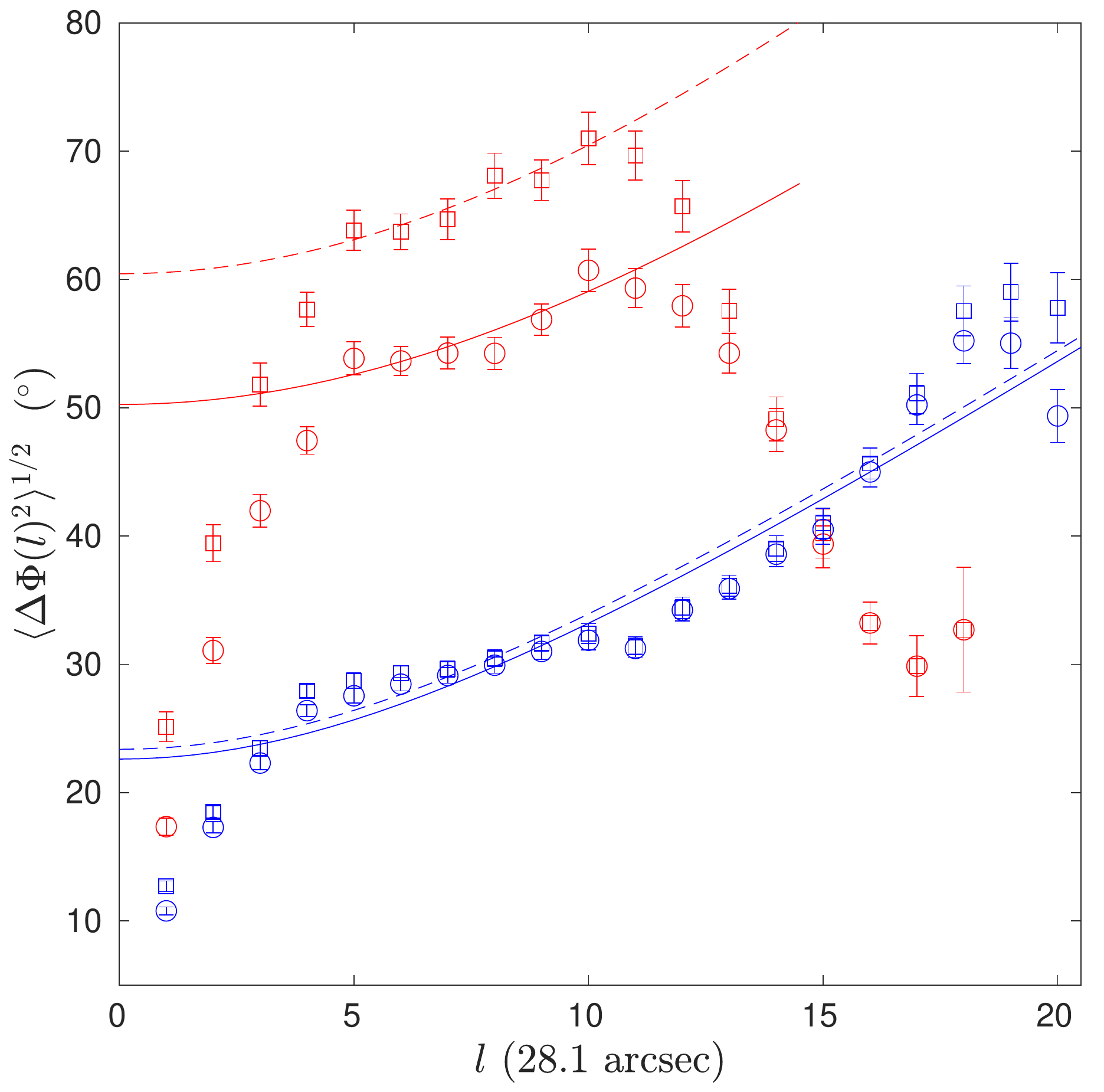}
\caption{Structure functions and fits for sub-regions SR1 (red) and SR2 (blue) using $\Delta\Phi$ restricted to less than 90$^\circ$ (circles and solid curve) and no restriction on $\Delta\Phi$ (squares and dash curve) as functions of scale in units of HAWC+ superpixels. For SR1, the fitting is from $\ell = 5$ to 10. The fitted intercepts with restriction and no restriction are $50.9\pm4.3^{\circ}$ and $60.9\pm3.0^{\circ}$, respectively. Error bars are the standard deviations of angle differences at a given distance. For SR2, the fitting is from $\ell = 4$ to 18. The fitted intercepts with restriction and no restriction are $23.2\pm3.5^{\circ}$ and $24.0\pm3.6^{\circ}$, respectively.
\label{fig:SF_fit}}
\end{figure}

While SR2 is dominated by one C$^{18}$O velocity component (\#12 with $V_{\rm LSR} =  5.6\,$km\,s$^{-1}$ in \citealp{hac13}), SR1 consists of two parts, SR1a to the north-east and SR1b to the south-west, where two distinct velocity components dominate (\#11 with $V_{\rm LSR} =  6.7\,$km\,s$^{-1}$ and \#12 with $V_{\rm LSR} =  5.6\,$km\,s$^{-1}$, respectively). These two components may be interacting with one another, increasing the dispersion in polarization angles. It may therefore seem justified to subdivide SR1 into these two parts (cf. SR1a and SR1b in Fig. \ref{fig:all_data}) when estimating the field strength with the DCF method. Doing this results in a polarization angle dispersion of $65^{\circ}\pm 9^{\circ}$ in SR1a for 45 independent HAWC$+$ measurements and a dispersion of $42^{\circ}\pm 5^{\circ}$ in SR1b for 73 independent measurements. In both parts of SR1, the dispersion of polarization angles remains significantly higher than in SR2. For both SR1 and SR2, we also analyzed the data using the structure function (SF) variant of the DCF method (\citealp{hil09}; section \ref{sec:DCF}, Appendix A)\footnote{We could not apply the SF technique to SR1a and SR1b separately, due to the low numbers of independent HAWC$+$ measurements in each of these smaller sub-regions.}. In Fig. \ref{fig:SF_fit}, we fit $\langle\Delta\Phi(\ell)^2\rangle^{1/2}$ for the two sub-regions using the SF method after correcting for measurement error by computing the error-weighted $\Delta \Phi^2(\ell)$ as in equation (\ref{eq:sigmatheta}). It is common practice to restrict $|\Delta\Phi|$ to be less than 90$^\circ$; that is, whenever $|\Delta\Phi|$ is found to be larger than 90$^\circ$, it is replaced by $|180^\circ-\Delta\Phi|$. As discussed in the Appendix, this often results in an underestimate of the dispersion in field angles and a corresponding overestimate of the field; in some cases, however, it can improve the accuracy of the field determination. We therefore provide both values. The intercepts of the fits at $\ell=0$ in Fig. \ref{fig:SF_fit} are $\Delta\Phi_0=50.3\pm4.4^{\circ}$ and $\Delta\Phi_0=60.5\pm3.0^{\circ}$ for the restricted and unrestricted approaches, respectively. The difference is only $10^{\circ}$. The corresponding angular dispersions contributed from the turbulence ($\sigma_{\theta} \sim \Delta\Phi_0/\sqrt{2}$) (eq. \ref{eq:sdb2}) are about $36^{\circ}$ and $43^{\circ}$, respectively. From the fitting, the turbulent correlation length scale of SR1 and SR2 is about 4 to 5 super-pixels, corresponding to 0.075 to 0.095 pc, about the size scale of the filamentary substructures. We divide the SR1 into four smaller portions, each with 30 polarization measurements. The root-mean-square of the angular dispersion in these four smaller portions is $30.3^{\circ}$, close to the angular dispersion of the turbulence from the SF analysis. The intercepts of fitting at $\ell=0$ in Fig. \ref{fig:SF_fit} are $23.2\pm3.5^{\circ}$ and $24.0\pm3.6^{\circ}$ from the restricted and un-restricted approaches, almost the same. The summary of all the measured parameters, and the results of the DCF and SF analysis are listed in Table \ref{tab:dcf_est}. The estimation for the ambient cloud around the entire L1495 is also provided in the table for comparison. 

\subsubsection{Volume density from Herschel column density data}
\label{sec:volume}
The average volume density in each of the two portions of the B211 filament marked in Fig.~\ref{fig:all_data} was estimated using the surface density map at 18.2$\arcsec $ resolution published by \citet{pal13} and \citet{mar16} from {\it Herschel} Gould Belt survey (HGBS) data\footnote{cf. \url{http://gouldbelt-herschel.cea.fr/archives}}. To do so, we assumed that the depth of each sub-region along the LOS is the same as the mean projected outer width. This is a very reasonable assumption, especially for SR2 which corresponds to a segment of the filament, since there is good observational evidence that B211 is a true cylinder-like filament as opposed to a sheet seen edge-on  \citep{ligold12}. The mean outer width was obtained using the projected area of pixels above the minimum surface density with a detected polarization signal [log $N({\rm H}_2) = 21.59$], divided by the length of each sub-region. The average surface density $< N({\rm H_2})>$  above log $N({\rm H}_2) = 21.59$ in each sub-region was derived from the {\it Herschel} column density map, and the resulting value was divided by the mean outer width, namely $L \sim 0.15$ pc for SR2 and $L \sim 0.3$ pc for SR1. This provided the average density, $<$$n({\rm H_2})$$>$\,$=$\,$<$$N({\rm H_2})$$>$/$L$, given in Table~\ref{tab:dcf_est}.

\subsubsection{Magnetic field strength}
\label{sec:bobs}

\begin{table*}
\caption{Summary of parameters and results of the DCF and SF analysis.}
\begin{tabular}{cccc}
\hline
\hline
Region\footnotemark & SR1 & SR2 & Taurus/B211\footnotemark\\
\hline
\vspace{0.2cm}
$N_{i}$\footnotemark & 120 & 162 & 175\\
\vspace{0.2cm}
$V_{\rm LSR}^{d}$\footnotemark - $V_{\rm LSR}^{m}$\footnotemark (km s$^{-1}$) & 5.4--5.9 & 5.5--5.5 & 6.6\\
\vspace{0.2cm}
$\sigma_{V}^{d}$\footnotemark - $\sigma_{V}^{m}$\footnotemark (km s$^{-1}$)& 0.26--0.48 & 0.27--0.41 & 0.85 $\pm$ 0.01\\
\vspace{0.2cm}
$\sigma_{\theta}$\footnotemark $(^{\circ})$ & 54 $\pm$ 5 & 20 $\pm$ 2 & 24 $\pm$ 2\\
\vspace{0.2cm}
$N(\rm H_2)$\footnotemark $(10^{21} {\rm cm}^{-2})$ & 8 $\pm$ 3 & 11 $\pm$ 4 & 1.5 $\pm$ 0.2\\
\vspace{0.2cm}
Depth\footnotemark (pc) & 0.3 & 0.15 & 0.5 $^{+0.5}_{-0.25}$\\
\vspace{0.1cm}
$n({\rm H_2})$\footnotemark $(10^4 {\rm cm}^{-3})$ & $1.0 \pm 0.4$ & $2.3\pm 1.0$ & 0.1$^{+0.1}_{-0.05}$\\
\hline
\vspace{0.2cm}
DCF analysis & & &\\
\vspace{0.2cm}
$B_{0}^{d}$\footnotemark - $B_{0}^{m}$\footnotemark (\mug) & 7--13 & 43--65 & 23$^{+12}_{-6}$\\
\vspace{0.2cm}
$\delta B^{d}$ - $\delta B^{m}$\footnotemark (\mug) & 10--18 & 16--24 & 10$^{+7}_{-5}$\\
\vspace{0.2cm}
$\muphi$ \footnotemark & 5.0--2.7
& 1.8--1.2 & 0.5 $^{+0.1}_{-0.1}$\\
\vspace{0.2cm}
$\ma/\cos\gamma$\footnotemark & 4.8  & 1.3  & 1.6 \\
\hline
\vspace{0.2cm}
DCF/SF analysis & & &\\
\vspace{0.2cm}
$\Delta\Phi_{0,\rm res}$\footnotemark/$\sqrt{2}$ $(^\circ)$ & $35 \pm 3$ & $16\pm 3$ & $14\pm 2$ \\
\vspace{0.2cm}
$\Delta\Phi_{0,\rm nores}$\footnotemark/$\sqrt{2}$ $(^\circ)$ & $42 \pm 2$ & $17\pm 3$ & $14\pm 2$ \\
\vspace{0.2cm}
$B_{0}^{m}$\footnotemark (\mug) & 17--23 & 79--82 & 41$^{+21}_{-11}$\\
\vspace{0.2cm}
$\delta B^{m}$\footnotemark (\mug) & 18 & 24 & 10\\
\vspace{0.2cm}
$\muphi$ & 2.5--2.1 & 1.0 & 0.3\\
\vspace{0.2cm}
$\ma/\cos\gamma$\footnotemark & 2.6--3.5 & 1.0 & 0.9\\
\hline
\label{tab:dcf_est}
\end{tabular}

\small
\begin{flushleft}
\fnt{1} {$^1$  Sub-region of B211 in which the analysis was conducted.}\\
\fnt{2} {$^2$ Estimation on a large scale covering the Taurus/B211 region using {\it Planck} polarization data \citep{pla13} and molecular line observations \citep{cha11}.}\\
\fnt{3} {$^3$ Number of independent SOFIA/HAWC$+$ polarization measurements for which $P/\sigma_{P}\geq 2 $, where $P$ is the polarized intensity. }\\
\fnt{4} {$^4$ Average centroid velocity of the dominant velocity component in each sub-region. }\\
\fnt{5} {$^5$ Average centroid velocity in each sub-region, including all velocity components.}\\
\fnt{6} {$^6$  Average non-thermal velocity dispersion of the dominant component over each sub-region.}\\
\fnt{7} {$^7$  Average value of the total non-thermal velocity dispersion over each sub-region.}\\
\fnt{8} {$^8$ Dispersion of polarization angles with individual measurements weighted by $1/\sigma_{{\theta}_{i}}^{2}$). The uncertainty in this dispersion was was estimated as $\sigma_{\theta}/\sqrt{N}$, where $N$ is the number of independent polarization measurements in each sub-region. \citep[cf.][]{pal20}}\\
\fnt{9} {$^9$ Weighted mean surface density derived from Herschel GBS data at the HAWC$+$ positions.}\\
\fnt{10} {$^{10}$ Adopted depth of each subregion estimated from the width measured in the plane of sky.}\\
\fnt{11} {$^{11}$ Average volume density estimated from $N(\rm H_2)$ and Depth.}\\
\fnt{12} {$^{12}$ Plane-of-sky mean field strength from the standard DCF method (eq.~\ref{eq:bpos2}) using the dispersion of the dominant velocity component.}\\
\fnt{13} {$^{13}$ Plane-of-sky mean field strength from the standard DCF method (eq.~\ref{eq:bpos2}) using the total non-thermal velocity dispersion.}\\
\fnt{14} {$^{14}$ The turbulent component of plane-of-sky $B$-field strength ($\delta B=B_0 \tan\sigma_{\theta}$, eq. \ref{eq:sdb0})}.\\
\fnt{15} {$^{15}$ Estimated mass to flux ratio relative to the critical value based on the rms POS field, $B_\tot=(B_0^2+\delta B^2)^{1/2}$ (eq. \ref{eq:muphipos}).}\\
\fnt{16} {$^{16}$ $\ma$ is the 3D \alfven Mach number ($\propto \surd 3 \sigma_V/B_{0,\rm3D}=\surd 3\sigma_V\cos\gamma/B_0$) with respect to the mean 3D field (eq. \ref{eq:ma}).}\\
\fnt{17} {$^{17}$ Intercept of the fitted structure function at $\ell=0$ with large-angle restriction.}\\
\fnt{18} {$^{18}$ Intercept of the fitted structure function at $\ell=0$ without large-angle restriction.}\\
\fnt{19} {$^{19}$ The range of $B_0$ estimated from $\Delta\Phi_{0,\rm nores}$ and $\Delta\Phi_{0,\rm res}$, respectively, using the total non-thermal velocity dispersion (eq. \ref{eq:bpos3}).}\\
\fnt{20} {$^{20}$ The turbulent component of plane-of-sky $B$-field strength 
$\delta B^m=\sdb=\fdcf(4\pi\rho)^{1/2}\sigma_V^m$ (eqs. \ref{eq:sdb2} and \ref{eq:bo4})}.\\
\fnt{21} {$^{21}$ $\ma$ is the 3D \alfven Mach number ($\propto \surd 3 \sigma_V/B_{0,\rm3D}=\surd 3\sigma_V\cos\gamma/B_0$) with respect to the mean 3D field} for the DCF/SF method (eq. \ref{eq:ma2}).\\
\end{flushleft}
\end{table*}

Using Equation~(\ref{eq:bpos2}) with the volume densities, velocity dispersions, and dispersions in polarization angles estimated in Sections 3.1.3 to 3.1.5, we can determine the field strengths for the two sub-regions marked by white dashed rectangles in Fig.~\ref{fig:all_data}. The results are summarized in Table \ref{tab:dcf_est}. We begin with SR2, which has a relatively smooth field with a small dispersion, $\sigma_\theta=20^\circ$. The field strength ranges between 43 $\mu$G and 66 $\mu$G from the standard DCF method and 79~\mug\ to 82~$\mu$G with the SF variant. Knowing the magnetic field, it is possible to determine the POS mass-to-flux ratio relative to the critical value, $\muphipos$, and the \alfven Mach number (Appendix \ref{app:equil}). SR2 is trans-\alfvenicstop, with $\ma\simeq 1.0-1.3$ and magnetically critical to mildly supercritical, $\muphipos\simeq 0.9-1.7$, depending on the method of analysis that is adopted. The critical mass per unit length, $M_{\crit,\ell}$, is that value of $M_\ell$ such that the pressure and magnetic forces are in balance with gravity (Appendix \ref{app:equil}). The SR2 filament segment is slightly subcritical, with $M_\ell\simeq 0.7 M_{\crit,\ell}$--i.e., it is gravitationally stable against radial collapse. In the absence of perpendicular magnetic fields, filaments that are moderately subcritical ($0.9\ga M_\ell/M_{\crit,\ell}\ga 0.2$, with an optimum value of $M_\ell/M_{\crit,\ell}\sim 0.5$) are subject to fragmentation into prestellar cores (i.e., starless cores with $M\geq\mbe$) since gas can flow along the filament (\citealp{nag87,fis12}; see Appendix \ref{app:equil}). Perpendicular fields suppress fragmentation for $\muphi< 1$. SR2 contains at least 5 candidate prestellar cores \citep{mar16}, which suggests that the lower estimates of the field in Table \ref{tab:dcf_est} are more accurate.

By contrast, SR1 has a chaotic field with a large dispersion in polarization angles, $\sigma_\theta =  54^{\circ}\pm 5^{\circ}$. This dispersion substantially exceeds the upper limit of applicability of the DCF method recommended by \citet{ost01}, as well as the less stringent criterion in Appendix {\ref{app:standard}. We note that the same remains true even if we subdivide SR1 into the two parts SR1a and SR1b considered in Sect.~\ref{sec:ang_dispersion}. Nonetheless, the large dispersion implies a small, albeit uncertain, field: The standard method yields $B_0\sim 7 - 13 ~\mu$G, depending on whether the velocity dispersion is estimated from the dominant velocity component ($B_0^d$) or the total line width ($B_0^m$). For the two sub-components of SR1, the DCF method gives $B_0\sim 10 - 11 ~\mu$G for SR1a and  $B_0\sim 22 - 28 ~\mu$G for SR1b. The large dispersion in angles is due in part to large scale variations in the field structure that are allowed for in the DCF/SF analysis. Using that method with the total line width, the estimated magnetic field strength $B_{0}^{m}$ is 23 or 16 $\mu$G, depending on whether $|\Delta\Phi|$ is restricted to be less than $90^\circ$ or not (see Appendix \ref{app:restrict}).  The turbulent magnetic field strength $\delta B_0^{m} = 18$ $\mu$G, comparable to $B_{0}^{m}$. We note that \citet{mar16} found only one candidate prestellar core in the sub-region SR1. Comparing the inset of figure 12 of \citet{hac13} with the Herschel column density map suggests that SR1 may be the location where material from the ambient cloud is presently being accreted onto B211. In particular, the fiber \#11 in \citet{hac13} is not straight and part of it is parallel to the striations seen in CO and Herschel data; it matches a “spur” or “strand” \citep[in the terminology of][]{cox16} and may correspond to the tip of a striation where it meets and interacts with the main B211 filament \citep{shi19}. This suggests that the flow velocities in the plane of the sky could be substantial, so that the observed LOS velocity is smaller than the POS velocities that determine $\sigma_\theta$. In fact, \citet{shi19} estimated that the inclination angle of the northeastern accretion flow to the line of sight is 70\degree, corresponding to a POS velocity 2.75 times larger than the LOS velocity. If so, the DCF value of the field there is an underestimate.

\citet{myer91} and \citet{hou09} have pointed out that if the turbulent correlation length, $\delta$, is less than the thickness of the region being observed along the LOS, $w$, then the dispersion in PAs will be reduced. \citet{hou09} found that the reduction factor is $[w/(2\pi)^{1/2}\delta]^{-1/2}$ when $\delta$ is much larger than the beam width. From Fig. \ref{fig:SF_fit} we find that $\delta$ is about 3 super pixels in size for SR2 and 5 super pixels for SR1, significantly greater than the beam width, which is less than one super pixel. In both cases, the turbulent correlation length is about $w/3$, so the reduction factor is of order unity. This is to be expected in a filament that forms in a turbulent medium. Since this effect is small compared to the uncertainties in the observations and in the method, we ignore it. 

It is instructive to compare DCF field measurements with Zeeman measurements. \citet{myer21} have applied the DCF method to a carefully selected set of low-mass cores and have shown that the measured magnetic fields give a median normalized mass-to-flux ratio, $\mu_{\Phi,\rm med}=1.7$, similar to that determined by the Zeeman method \citep{crut10}. As they note, there are very few cores with both DCF and Zeeman measurements. There are no Zeeman measurements of the field in B211, so we compare with the average Zeeman field in interstellar molecular clumps determined by \citet{li15} from the Zeeman data summarized by \citet{crut10}. The average LOS field is $B_{\rm Zeeman,\, LOS}=33 n_{\htwom,4}^{0.65}$~\mug. The median angle of inclination between a filament and the plane of the sky is $\gamma=30^\circ$, so the mean POS field inferred from the LOS field is $(\tan 30^\circ/\tan \gamma)B_{\rm LOS}$. This is the mean field, not the total or rms field, since that is what Zeeman observations measure. The POS field corresponding to the average Zeeman field is thus
\beq
B_{{0,\rm Zeeman}}=57 \left(\frac{\tan 30^\circ}{\tan \gamma}\right)n_{\htwom,4}^{0.65}~~~\mmug.
\eeq
For SR1, with $n_{\htwom,4}=1$, this gives an inferred POS field (not a measured one) of 57~\mug\ at the average inclination, much larger than the 13-23~\mug\ from the DCF methods with the full line profile. This suggests that the DCF method indeed underestimates the field in this region. For SR2, with $n_{\htwom,4}=2.3$, the inferred Zeeman POS field is 98~\mug, a little larger than the DCF estimates, 66-82~\mug. 

We also estimated the field strength of a larger area of Taurus/B211 using {\it Planck} polarization data from \citet{pla13} (at an effective HPBW resolution of 10$\arcmin$). The independent polarization measurements from {\it Planck} in this area (displayed as orange vectors in Fig.~\ref{fig:all_data}) indicate a dispersion in polarization angles of about 24$^\circ $ at 10$\arcmin$ resolution. The average velocity dispersion in this extended environment around almost the entire L1495/B213 filament is $\sim$0.85~km~s$^{-1}$ as estimated by \citet{cha11} from $^{13}$CO(1--0) observations. We estimated the average volume density, $n_\htwom\simeq 10^3$~cm\eee, following the same approach as described in Section 3.1.5 but adopting a characteristic depth of $\sim$\,0.5 pc for the ambient cloud around Taurus/B211 \citep[see][]{shi19}. Applying the DCF formula of Equation~(\ref{eq:bpos2}) with these values lead to a field strength of $\sim 41\, \mu$G.

\subsection{Polarization vectors and surface density contours}
\label{sec:polden}

As discussed in \citet{sol17} and references therein, the gas that feeds a cloud appears to be gathered along the magnetic field direction. Physically, it is easier for gas to flow along the field than perpendicular to the field when the field is dynamically important. Furthermore, a long, slender filament can accrete gas much more easily on its sides than at its ends. This accounts for the observation that the dense regions in many molecular clouds show magnetic fields that tend to be perpendicular to contours of the surface density \citep{pla16}.

\begin{figure*}
\includegraphics[scale=0.73]{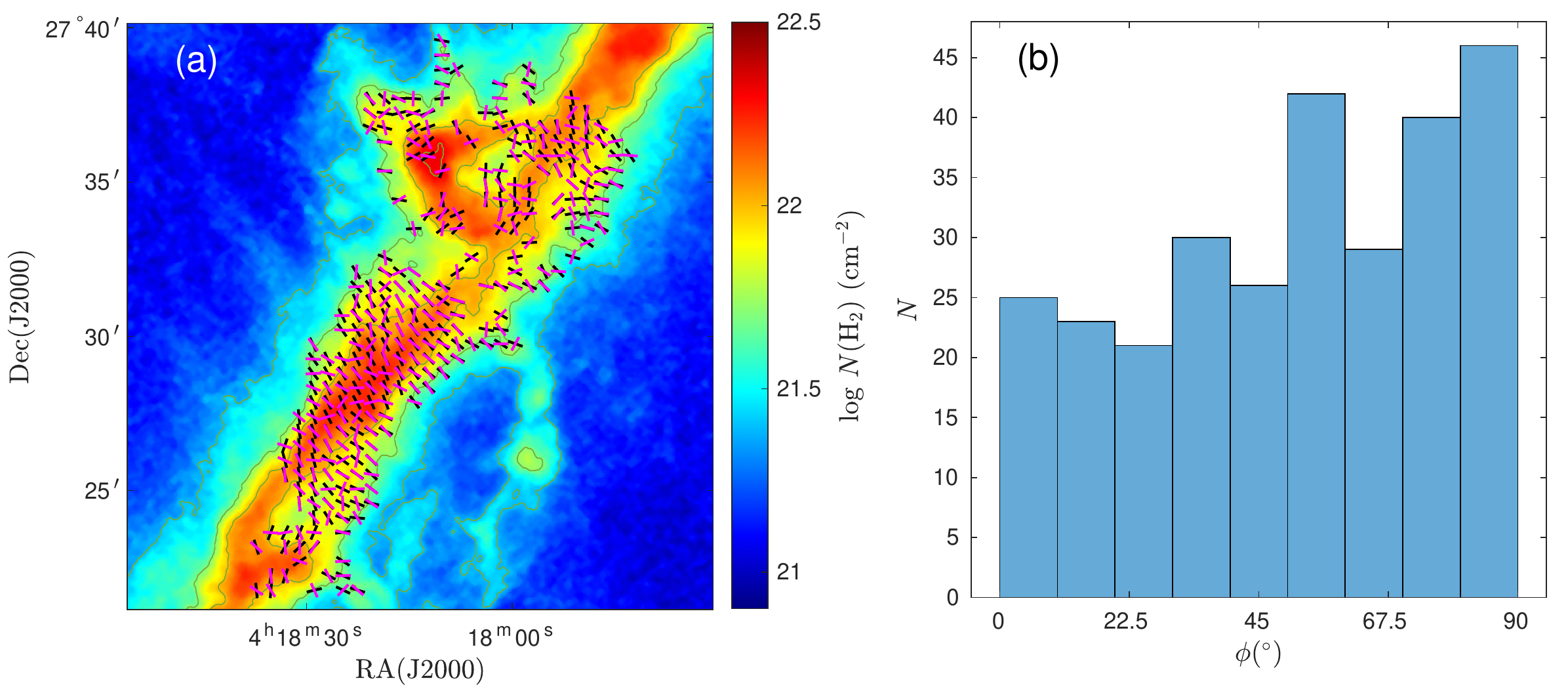}
\caption{(a) Orientations of the inferred magnetic field (black) from HAWC+ and of the surface density gradient vectors (magenta) are plotted over the Herschel surface density. (b) Histogram of the distribution of angles between the inferred magnetic field directions and the tangents of the contours of surface density.
\label{fig:ngrad_b}}
\end{figure*}

Let $\phi$ be the magnitude of the angle between the field vector inferred from polarization (i.e., the PA) and the tangent to the surface density contour, so that $0^\circ\leq\phi\leq 90^\circ$. \citet{sol13} found that in many of the cases they considered the magnetic field tended to be parallel to the isodensity contours in 3D and column density contours in 2D ($\phi\sim 0^\circ$). For strong fields ($\beta=8\pi\rho\cs^2/B^2\la 1$), the relative orientation became closer to perpendicular ($\phi\sim 90^\circ$) at high densities. \citet{sei20} attributed the change in relative orientation at high density to the gravitational energy becoming comparable to the magnetic energy. An alternative description of the $\phi$ distribution was introduced by \citet{sol13,sol17}, the histogram shape parameter:
\beq
\xi = \frac{A_0-A_{90}}{A_0+A_{90}},
\label{eq:eta}
\eeq
where $A_0$ is the area under the histogram of $\phi$ values for 0$^\circ\leq\phi\leq 22.5^\circ$ and $A_{90}$ is the area for 67.5$^\circ\leq\phi\leq 90^\circ$. A negative value of $\xi$ means that the PAs tend to be perpendicular rather than parallel to the surface contours. The ratio of perpendicular to parallel PAs is $A_{90}/A_0=(1-\xi)/(1+\xi)$.

Gas flows near the cloud determine how gas is accreted onto the cloud and thus how the cloud forms \citep{shi19}. However, observations provide only the LOS velocity information, which can be very different from the POS velocity and thereby give a misleading idea of the true spatial gas movement \citep{li19}. As noted above, fields with a substantial component normal to a filament can facilitate accretion of gas onto the filament. To assess the importance of magnetic fields in B211, we present two complementary plots of the data. In Fig. \ref{fig:ngrad_b}a, we plot the orientations of the gradient of the surface density from \textit{Herschel} data against the PAs from the HAWC+ observations. Note that the gradient of the surface density is normal to the contours of surface density, so that fields perpendicular to the filament are parallel to the gradient. In Fig. \ref{fig:ngrad_b}b, we plot a histogram of the $\phi$ distribution (i.e., the distribution of angles between the PAs and the tangents of the surface density contours). Because of the relatively small number of detected pixels and the limited dynamic range of the SOFIA polarization data in terms of column density, we cannot meaningfully apply a tool such as the histogram of relative orientations (HROs) as a function of surface density to the HAWC+ observations in B211. Therefore, we show only one HRO in Fig. \ref{fig:ngrad_b}b from all the detected pixels of the observed B211 region. The histogram shape parameter for B211 is $\xi=-0.28$. The negative value is primarily due to SR2, which has $\xi=-0.48$; the chaotic field in SR1 has $\xi\sim 0$. It is clear from Fig. \ref{fig:ngrad_b}b that there are more pixels at $90^{\circ}$ than at $0^{\circ}$. The distribution of angles in this figure is similar to the high surface density Centre-Ridge region in the Vela C molecular complex. As noted above, a negative value of $\xi$ is consistent with gas accretion along field lines that thread the cloud.  

\section{Comparison with simulation}
\label{sec:sim_comp}
Above we used observational data from HAWC+ and the IRAM 30m telescope to obtain the LOS velocity, the magnetic field orientation, and an estimate of the field strength. In this section we shall compare these observations with a numerical simulation that was designed not to simulate L1495 in particular, but rather to simulate the formation of filamentary structures in a typical supersonically turbulent, magnetized interstellar molecular cloud \citep{li19}. Although there are some differences between the simulated filamentary cloud and L1495, such as the mass per length and probably the overall magnetic field strength in the regions, the filamentary substructures in the simulated cloud are similar to those in L1495 \citep{hac13}. In fact, the results of our simulation inspired this high-resolution polarization observation of the L1495/B211 region with the aim of understanding the three-dimensional structure of the magnetic field inside filamentary clouds. 
\begin{figure*}
\includegraphics[scale=0.6]{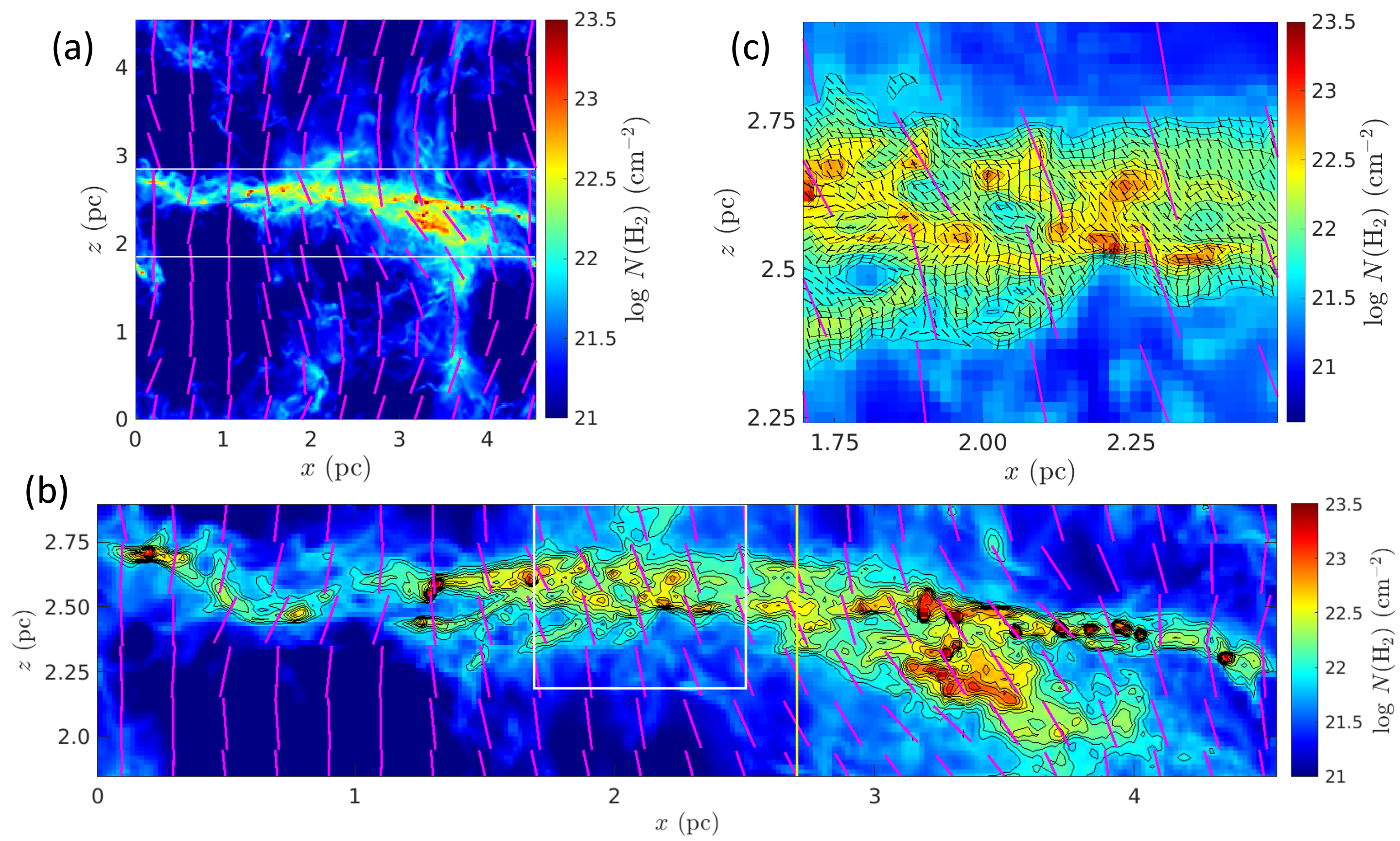}
\caption{(a) Surface density map of the entire simulated region as viewed along the $y$-axis; the mean field is in the $z$ direction. Magenta lines indicate the large-scale magnetic field at a resolution of 0.4 pc. The filamentary cloud lies between the two long white lines. (b) Enlargement of the surface density map of the filamentary cloud with contours of log$_{10}N({\rm H_2})$ ranging from 21.875 to 23, separated by $\Delta \log_{10}N({\rm H_2})=0.125$. Due to a collision of two filamentary clouds near $x = 3.3$ pc, only the range $x = 0$ to 2.7 pc (up to the yellow vertical line) of the cloud is used for comparison with observation. Magenta lines indicate the large-scale magnetic field at 0.2 pc resolution.  (c) Zoom in around the 0.82 pc $\times$ 0.69 pc FOV window in panel (b) showing the highly perturbed magnetic field at $28\farcs1$ resolution (the same as the HAWC+ observation). The local orientation of the field is indicated by the short black lines (shown only at pixels with log$_{10}N({\rm H_2}) \geq 21.59$). As in (b), magenta lines indicate the large-scale magnetic field at 0.2 pc resolution. The surface density contours start from log$_{10}N({\rm H_2})$ of 21.59 and separated by $\Delta \log_{10}N({\rm H_2})=0.125$.
\label{fig:simcloud}}
\end{figure*}

To compare the HAWC+ observational results in Section \ref{sec:hawc} with simulation, we use our high-resolution simulation results of the formation of filamentary molecular clouds described in detail in \citet{li19}. This simulation used our multi-physics, adaptive mesh refinement (AMR) code \textsc{Orion2} \citep{li12}. Since the purpose of the simulation was to study the formation of filamentary structures prior to the onset of star formation, radiation transport and feedback physics were ignored. The ideal MHD simulation begins with turbulent driving but without gravity for two crossing times in order to reach a turbulent equilibrium state. The entire simulation region is 4.55 pc in size with a base grid of $512^3$. Two levels of refinement were imposed to refine pressure jumps, density jumps, and shear flows to reach a maximum resolution of $2.2 \times 10^{-3}$ pc, which was chosen to be sufficient to study filamentary substructures with a width of order 0.1 pc. Turbulence was driven throughout the simulation at a 3D thermal Mach number $\calm = 10$ on the largest scales, with wave number $k = 1 - 2$. Gravity was turned on after two crossing times. After gravity was turned on, we included an additional refinement requirement, the Jeans condition \citep{tru97}. We adopted a Jeans number of 1/8, which means that the Jeans length is resolved by at least 8 cells. We adopted periodic boundary conditions and assumed an isothermal equation of state for the entire simulation at a temperature of 10K. Using the turbulent line-width-size relation \citep{mck07}, setting the \alfven Mach number to be 1, and setting the virial parameter to be 1, implies that the total mass of the entire cloud is $M = 3110 \,M_{\sun}$ and the initial magnetic field is 31.6 \mug. A long, massive filamentary cloud formed after gravity was turned on, and at a time of 700,000 yr, it had a length of 4.42 pc and a mass of about $471 \,M_{\sun}$. The moderately strong large-scale field was found to be crucial in maintaining the integrity of the long and slender filamentary cloud. Details of the physical properties of the filamentary cloud can be found in \citet{li19}.

\subsection{Simulation parameters and methods}
\label{sec:sim}
\begin{figure*}
\includegraphics[scale=0.78]{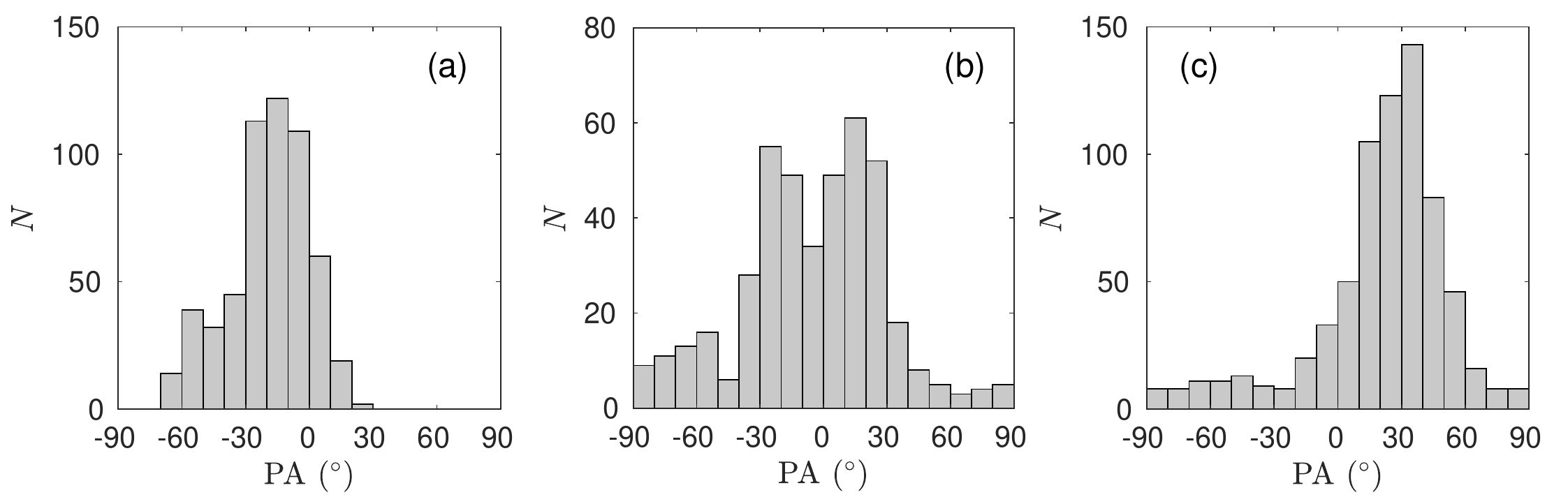}
\caption{PA distributions of magnetic field of three 0.82 pc long FOVs of the simulated cloud. The angle $\theta$ is measured relative to the mean direction of the field in the simulation box. (a) Single group PA distribution from a FOV starting at $x$ = 0.16 pc; the distribution peaks at $-25^{\circ}$. (b) Double-hump PA distribution from a FOV starting at $x$ = 0.75 pc. (c) A single group PA distribution from a FOV starting at $x$ = 1.69 pc, with a peak at $35^{\circ}$, at the location of the FOV window in Figure \ref{fig:simcloud}b.
\label{fig:pasim}}
\end{figure*}

In our simulation, even the base grid has resolution of $\sim 0.009$ pc per cell, higher than HAWC+ superpixels. To produce the same resolution map for direct comparison with HAWC+ or Planck data, we first integrate LOS quantities, such as volume density to obtain the surface density, over the base grid to create a 2D map at $512^2$ resolution. We compute the Stokes parameters following \citet{zwe96}. Density weighting is used when computing the Stokes parameters and the LOS velocity dispersion. We then coarsen the 2D map to the resolution of a HAWC+ superpixel or of the Planck data by computing the mean of the corresponding number of pixels.

The surface density map of the entire simulated region is shown in Fig. \ref{fig:simcloud}a. The polarization field indicating the density-weighted large-scale magnetic field at a resolution of 0.4 pc, which is the best resolution that Planck can achieve at the distance of L1495, is superimposed on the map. In the other two panels of Fig. \ref{fig:simcloud}, the large-scale polarization field are shown at 0.2 pc resolution. The main filamentary cloud in between the two white lines is enlarged in Fig. \ref{fig:simcloud}b. The simulated filamentary cloud is composed of rich filamentary substructures along the entire length, similar to L1495 and other filamentary clouds. To study the magnetic field structures of filamentary clouds at the early stage of the formation, it is helpful to observe a cloud before the formation of protostars because powerful protostellar outflows can disrupt the magnetic field structures within filamentary substructures. The region B211 in L1495 has no protostars but contains filamentary substructures \citep{hac13} and prestellar cores \citep{mar16}. Therefore, the simulated cloud is suitable for comparison with B211. Due to the collision of two filamentary clouds in our simulation at $x \sim 3.3$ pc, our comparison with observations will be in the range of $x = 0 - 2.7$ pc, i.e. up to the left of the vertical yellow line in Fig. \ref{fig:simcloud}b. The length of B211 with signal detected by HAWC+ is about 0.82 pc. We can create a projection of the cloud of the same length within this range for comparison. An example of a projected window, the white box in Fig. \ref{fig:simcloud}b, is shown in Fig. \ref{fig:simcloud}c. The small scale magnetic field structures at the resolution of 0.019 pc, corresponding to the super-pixel resolution in the HAWC+ observation, are shown together with the low resolution magnetic field. All the following comparisons between the simulation and HAWC+ observations will be at this resolution. For clarity, we show only vectors at pixels with surface density log$_{10} N({\rm H}_2) \ge 21.59$, corresponding to the minimum surface density with detected polarization signal in the observed B211 region by HAWC+. We can see the small-scale magnetic fields inside the cloud have large deviations from the low resolution large-scale fields surrounding the dense substructures, as shown in \citet{li19}.

\subsection{PA distribution}
\label{sec:pa}

The HAWC+ observations of B211 show a larger dispersion of PAs than the lower resolution Planck observations of the large-scale field as discussed in Section \ref{sec:polarization}. The results indicate that small-scale perturbations of the magnetic field are present in B211. In Fig. \ref{fig:pasim}, we show the PA distributions of three FOVs in the simulation. They have a length of 0.82 pc, which corresponds the HAWC+ map, and height of 0.69 pc, which is large enough to include the width of the filament. The distribution in Fig. \ref{fig:pasim}a is a single group peaking at about $-15^\circ$. In Fig. \ref{fig:pasim}b, the distribution becomes double-humped, with peaks at $-25^\circ$ and $15^\circ$. These two FOVs along the filamentary cloud have quite different PA distributions even though they are offset by only 0.6 pc. In Fig. \ref{fig:pasim}c, which is the white colored FOV shown in Fig. \ref{fig:simcloud}b, the distribution returns to a single group again and peaks near $35^\circ$. We see that the PA distribution and the mean PA vary along the simulated cloud. In \citet{pal13}, the mean PA of the extended optical and infrared polarization vectors also changes along the filamentary cloud L1495. More polarization mapping in different parts of the filamentary cloud Taurus/B211 will be needed to find out if the PA distribution would change as in Fig. \ref{fig:pasim}.

\begin{figure}
\includegraphics[scale=0.35]{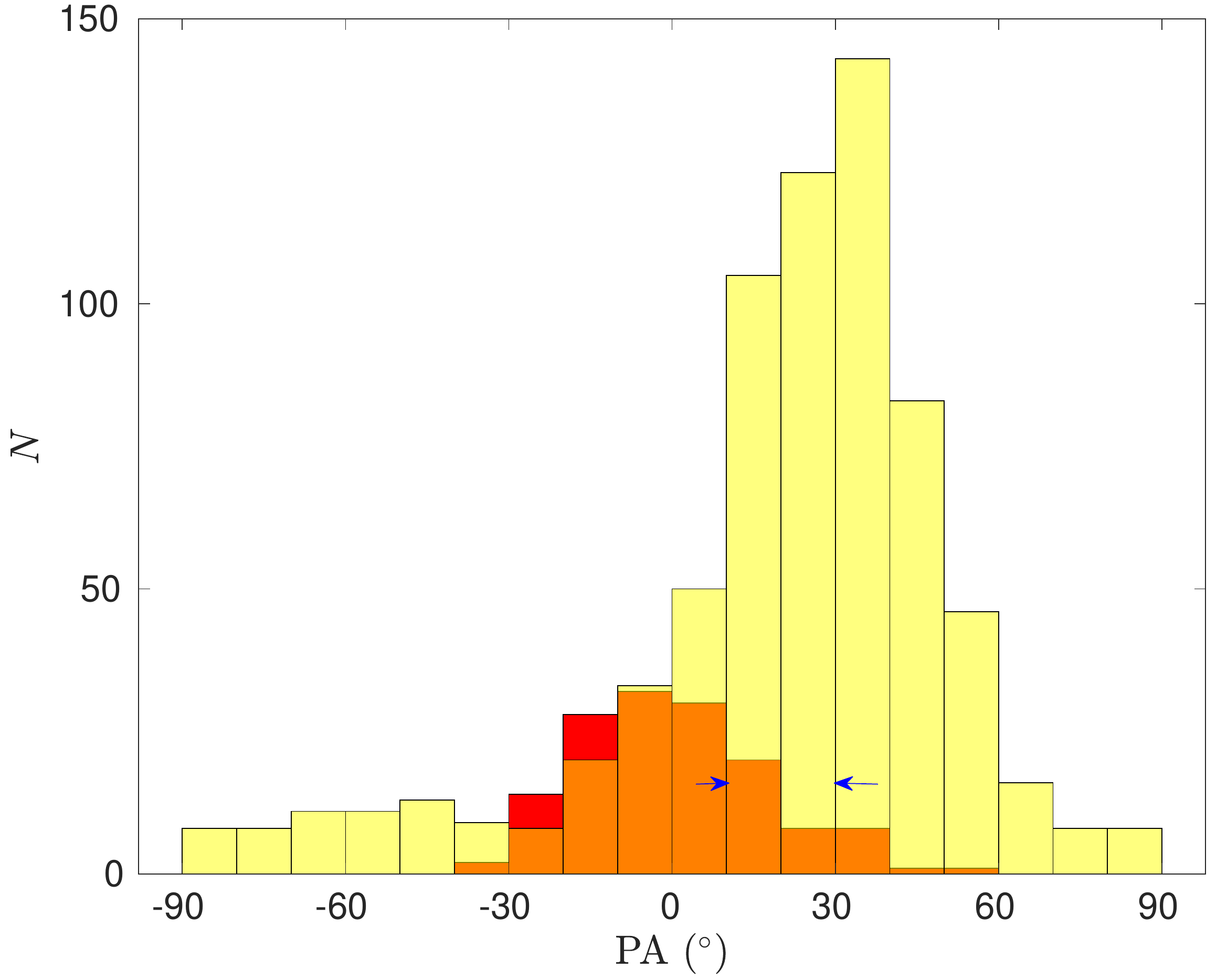}
\caption{Comparison of the PA distributions of magnetic field in the simulation FOV window of Fig. \ref{fig:simcloud}b at HAWC+ 28.1 arcsec resolution (yellow histogram) and the entire simulation box (Fig. \ref{fig:simcloud}a) at Planck 10 arcmin resolution (red histogram; overlapping points are in orange). The 10 arcmin low resolution field vectors inside the simulation FOV window are all within the two bins between $10^\circ - 30^\circ$, marked by the blue arrows.
\label{fig:PA_resolcomp}}
\end{figure}

In Fig. \ref{fig:PA_resolcomp}, we compare two PA distributions in the simulation by viewing the simulated cloud at the same distance of L1495, one in a small region at the HAWC+ superpixel resolution of 28.1 arcsec and one in the whole simulated box at the Planck resolution of 10 arcmin. For the small region, we choose the FOV outlined in Fig. \ref{fig:simcloud} since the PA distribution of this segment of the simulated cloud is similar to that of the observed B211 region. The other two FOV windows are quite different from B211, so we shall not discuss them further. The PA distribution at the Planck 10 arcmin resolution (red histogram in Fig. \ref{fig:PA_resolcomp}) is obtained from all the vectors in Fig. \ref{fig:simcloud}a. At this resolution the dispersion is only $6.6^\circ$, much smaller than the dispersion of $29.2^\circ$ of the polarization at the HAWC+ superpixel scale (see Table \ref{tab:dcf_sf_sim}). The resolution effect on the dispersion of PAs is clear both in simulation and observation (Fig. \ref{fig:observed_pa}a). Some of this reduction in dispersion is likely due to a much lower dispersion in the low column-density gas that fills much of the Planck field: We found a dispersion of only 6.8\degree\ in a low-column region above the FOV window in the simulation. Since the polarization in the high-column LOSs is dominated by emission in the filament whereas that in the low-column LOSs is spread more uniformly over the entire LOS, the dispersion in the low-column directions is reduced by averaging along the LOS. In other words, the longer effective path length in the low-column directions leads to an LOS resolution effect.

In addition to this observational effect, the dispersion of PAs inside a molecular cloud is increased by the combined results of differential motions of dense substructures during cloud formation \citep{li19} and small-scale local gravity-driven motion as seen in numerical simulations of molecular cloud formation  \citep[e.g][]{che16,li19,sei20}. These motions stretch the magnetic field locally, causing large changes in the direction of the magnetic field, as shown in figure 10 of \citet{li19} and Fig. 9c in this paper.

\subsection{DCF field estimates in the simulation}
\label{sec:dcf_sim}

\begin{table}
\caption{Comparison of physical properties of the simulated filamentary cloud estimated from DCF and DCF/SF methods at HAWC+ resolution}
\label{tab:dcf_sf_sim}
\begin{tabular}{lcc}
\vspace{-0.3cm}\\
\hline
\hline
Method & DCF & DCF/SF \\
\hline
$w ~~ ({\rm pc})$ & 0.33 & 0.33 \\
$n({\rm H_2}) ~~({\rm cm}^{-3})$ & $1.52\times10^4$ & $1.52\times10^4$ \\
$\sigma_{\theta} ~~ (^\circ)$ & 29.2 & - \\
$\Delta\Phi_{0,\rm res} ~~ (^\circ)$ & - & 35.4 \\
$\Delta\Phi_{0,\rm nores} ~~ (^\circ)$ & - & 37.2\\
$\sigma_V ~~({\rm km~s}^{-1})$ & 0.45 & 0.45 \\
$B_{0,\rm DCF} ~~({\rm \mu G})$ & 38.0 & 41.0 - 43.6\footnotemark \\
$\delta B_{\rm DCF} ~~({\rm \mu G})$ & 21.2 & 21.2 \\
$B_{0,\rm true} ~~({\rm \mu G})$ & 55.9 & 55.9 \\
$\delta B_{\rm true} ~~({\rm \mu G})$ & 55.9 & 55.9 \\
\hline
\end{tabular}
\begin{flushleft}
\fnt{1} {$^1$ The smaller value is obtained using $\Delta\Phi_{0,\rm nores}$, the value obtained without restricting $\Delta\Phi$ to be in the range $0^\circ-90^\circ$.}\\
\end{flushleft}
\end{table}

Here we apply the DCF and DCF/SF methods at HAWC+ super-pixel resolution to the FOV outlined in white in Fig. \ref{fig:simcloud}b. The velocity dispersion is density-weighted along the LOS. The mean width of the simulated filament, $w=0.33$~pc, was computed by dividing the projected area of pixels above log $N({\rm H}_2) = 21.59$ (the minimum surface density of the observed B211 region with a detected polarization signal) by the 0.82 pc length of the segment. Following the procedure used in analyzing
the observations, we then estimated the density from the column density by assuming that
the mean depth of the cloud is the same as the mean width, $w$.

In Table \ref{tab:dcf_sf_sim}, we compare the results for the simulated cloud at HAWC+ super-pixel resolution using the DCF and the DCF/SF methods. The turbulent correlation length, $\delta$, in the simulated cloud segment is about 5 to 6 super-pixels, similar to that in SR1 and SR2 (see Fig. \ref{fig:A21}). This length is resolved by more than 40 cells at the highest resolution, so the DCF/SF results should be reliable. The magnetic field strength estimated using the DCF/SF method is in the range $41.0 - 43.6$ \mug, a little larger than the estimated value using DCF method and slightly closer to the true value.

The mean volume density, $n({\rm H_2})$, LOS velocity dispersion, $\sigma_V$, and dispersion of polarization PA, $\sigma_{\theta}$, are listed in Table \ref{tab:dcf_sf_sim}, and the  \alfven Mach number based on the POS field in this window, $\ma/\cos\gamma$, is given in Table \ref{tab:properties}; all these values are intermediate between the values for SR1 and SR2 given in Table \ref{tab:dcf_est}. The primary difference between the simulated and observed regions is that the simulation has a higher mass per unit length, $M_\ell$, and correspondingly a smaller value of the filament virial parameter, $\avir=2\sigma_V^2/(GM_\ell)$ \citep{fie00a} (Table \ref{tab:properties}); the magnetic properties are similar.

To determine the true mean POS magnetic field strength above the surface density threshold in the simulation, the volume-means of the two projected magnetic field vector components for each pixel above the threshold were computed first. These vector fields were then averaged along the line of sight to a depth of 0.8 pc to obtain the mean vector field, $\vecB_{0,\rm true}$. Since the value of the true field is volume weighted whereas the DCF field is based on the density-weighted polarization (which is quite different from a density-weighted field), we do not expect the true field to exactly agree with the DCF field. The true 3D field is 57.1 \mug, which implies that the mean field is at an angle of 12\degree\ with respect to the plane of the sky. Table \ref{tab:dcf_sf_sim} shows that the DCF and DCF/SF estimates of the mean POS field are about 70 percent of the true value. The SF variant is slightly closer to the true value than the standard DCF method, and the restricted value for the SF variant is slightly closer than the unrestricted value, but given the uncertainty in $\fdcf$ and the fact that the true field and the DCF field have different weightings, it is not clear that these differences are significant.

The turbulent field, $\delta B_{0,\rm true}$, is the root-mean-square of the vector difference of the field vectors from the volume-mean field vector in all the cells in the volume corresponding to the FOV. The true values of $\delta B$ are significantly greater than the true values of $\delta B_\perp$ (the rms value of component $\delta\vecB$ perpendicular to the mean POS field) due to the substantial parallel component of $\delta\vecB$ in $\delta B$. The value of $\delta B_\perp$ is the same for the DCF and DCF/SF methods since both are based on equation (\ref{eq:sigmav}). The approximations made in the DCF/SF method imply that the value of $\delta B$ calculated in this method is actually $\delta B_\perp$. The fact that the true value of $\delta B_\perp$ is about 1.7 times larger than the DCF value, $B_0\tan\sigma_\theta$, accounts for most of the difference between the DCF value of $B_0$ and the true value.

The derived physical parameters of the observed sub-regions in B211 and the segment of the simulated cloud are summarized in Table \ref{tab:properties}. The values of $\avir$ are based on the total velocity dispersions (including thermal motions) of SR1, SR2, and the segment of the simulated cloud. The normalized mass-to-flux ratio in the simulation based on the DCF field is $\muphipos=2.7$, comparable to the observed values in SR1. Since the true value of the total POS field is 79~\mug, the true value of $\muphipos$ is 1.5, intermediate between the values for SR1 and SR2 and comparable to the initial value of 1.62 of the entire box viewed normal to the initial field. It is this value that we have used in determining $\muphipos$ in the simulation. (Note that the use of the POS field to estimate $\muphi$ leads to a slight overestimate of $M_\ell/M_{\rm crit,\ell}$--see eq. \ref{eq:cositot}.) Including the effect of a perpendicular magnetic field, the ratio of the line mass to the critical line mass, $M_\ell/M_{\rm crit,\ell}$ (eq. \ref{eq:mellcr}), is less than unity for both observed subregions and  for the simulated cloud segment, as shown in the table. These structures are therefore gravitationally stable against radial collapse on the scale at which this ratio is determined. In the absence of perpendicular magnetic fields, the critical mass and the virial mass are the same, and the filaments would be subject to fragmentation (\citealp{nag87,fis12}; see Appendix \ref{app:frag}). Perpendicular fields would stabilize the cloud against fragmentation if the normalized mass-to-flux ratio is $\muphi< 1$. The results in Table \ref{tab:properties} show that SR1 and the simulated filament should be subject to fragmentation; since SR2 has $\muphi\sim 1$ it is marginally susceptible. There is some evidence that pre-stellar cores are forming in both SR1 and SR2 \citep[e.g.,][]{mar16}; as noted above, the fact that pre-stellar cores are observed in SR2 favors a lower estimate for the field there than the value given in Table \ref{tab:properties}. Equation \ref{eq:cf} shows that fragments that form in SR1 are near the critical mass and could collapse, whereas those in SR2 appear to be stable against collapse; however, given the uncertainty in the parameters in Table \ref{tab:properties}, these conclusions are tentative. The FOV window in the simulation is magnetically supercritical, and dense cores are forming along some filamentary substructures. On small scales, we expect the velocity dispersion to be primarily thermal. At a temperature of 10~K, the thermal values of the virial parameter are $\avir= 0.30,\, 0.46,\, \mbox{and}\; 0.15$, respectively.

\begin{table}
\caption{Summary of physical properties of the observed B211 sub-regions and a segment of the simulated filamentary cloud.$^1$}
\label{tab:properties}
\begin{tabular}{lccc}
\vspace{-0.3cm}\\
\hline
\hline
Region & SR1 & SR2 & simulation \\
\hline
$M_\ell (M_{\odot}\; {\rm pc}^{-1})$ & 54 & 36 & 111 \\
$\avir$\footnotemark & 2.0 & 2.2 & 0.85 \\
$B_{0,\rm DCF}$ (\mug) & 13 - 23\footnotemark & 65 - 82 & 38 - 44 \\
$B_{\tot,\,\rm DCF}$ (\mug)\footnotemark & 23 - 30 & 70 - 85 & 44 - 48 \\
$\mu_{\Phi,\,\rm DCF}$\footnotemark & 2.7 - 2.1 & 1.2 - 1.0 & 2.7 - 2.4 \\
$\ma/\cos\gamma$\footnotemark & 4.8 - 2.6 & 1.3 - 1.0 &  2.0 - 1.6 \\
$M_\ell/M_{\rm crit,\ell}$\footnotemark & 0.50-0.49 & 0.43 - 0.42 & 0.9$^7$ \\
\hline
\end{tabular}
\begin{flushleft}
\fnt{1} {$^1$ The virial parameter for a filament is $\avir= 2 \sigma_V^2/(G M_\ell)$.}\\
\fnt{2} {$^2$ The two values quoted for parameters that depend on the magnetic field correspond to the DCF and the larger of the two DCF/SF estimates, respectively, of the field strength using the total non-thermal velocity dispersion.}\\
\fnt{3} {$^3$ The total DCF field, $B_{\tot,\,\rm DCF}=(B_{0,\rm DCF}^2+\delta B_{\rm DCF}^2)^{1/2}$.}\\
\fnt{4} {$^4$ Normalized mass-to-flux ratio based on the total DCF field,
$B_{\tot,\,\rm DCF}$. The value based on the true total field is $\mu_{\Phi,\,\rm true}=1.5$.}\\
\fnt{5} {$^5$ 3D \alfven Mach number ($\propto \surd 3\sigma_V$) based on the mean POS field, $B_{0,\rm DCF}$, and assuming isotropic turbulence (eq. \ref{eq:ma}).}\\
\fnt{6} {$^6$ $M_\ell$ is the mass per unit length for $N(\htwom)\geq 10^{21.59}$ cm\ee. The critical line mass, $M_{\ell,\,\crit}$, is given in equation (\ref{eq:mellcr}).}\\
\fnt{7} {$^7$ Based on the true value of $\muphipos=1.5$.}
\end{flushleft}
\end{table}

\section{Conclusions}
In this work, we have used HAWC+ on-board SOFIA to observe the L1495/B211 region in Taurus to investigate the magnetic field morphology in thin filamentary clouds. This observation is challenging because of the low surface brightness of the filamentary cloud. We needed to re-sample $3 \times 3$ detector pixel data to a super-pixel of $28\farcs1$ to optimize the SNR. We have total 282 independent measurements that have $P/\sigma_P \ge 2$. The morphology of the observed polarization map clearly reveals two sub-regions, designated SR1 and SR2, in the observed B211 region. With IRAM 30m C$^{18}$O (1-0) data, we estimate the magnetic field strengths using the standard DCF method and the alternative DCF method using a structure function. We then compared the physical states of the two sub-regions with a simulated filamentary cloud.

\begin{itemize}
\item[1.] {\it Polarization morphology of the two sub-regions in B211.}
The chaotic appearance of the polarization vectors in SR1 indicates a strongly perturbed region, in contrast to SR2, which has a well organized magnetic field structure mostly, but not entirely, perpendicular to the filamentary cloud axis. The organized field in SR2 matches the large-scale field from Planck observation very well. The dispersion of the PAs in SR1 is $54\deg$, almost 3 times of that in SR2.

\item[2.] {\it Filamentary substructures in B211.}
The IRAM 30m C$^{18}$O (1-0) data reveals multiple velocity components in the observed B211 region, similar to what has been reported in a previous study of L1495 \citep[e.g.][]{hac13}. There are at least 3 velocity components in SR1 and 2 velocity components in SR2. Multiple filamentary substructures are also clearly seen in the high resolution Herschel map. The chaotic appearance of substructures, polarization vectors, and the multiple-component line profiles in SR1 may indicate strong interaction among substructures.

\item[3.] {\it Magnetic fields of the two sub-regions and of the simulated filament.}
Using the DCF and DCF/SF methods, the estimated field strength based on the total LOS velocity dispersion in SR1 is in the range 13 to 23 \mug. Because of the very large dispersion of the polarization angles ($\sigma_\theta=54^\circ$), the field estimate in this region is very uncertain, but it is clear that the field is small. By contrast, the estimated field strength of SR2 is from 66 to 82 \mug, significantly larger than that in SR1. These estimates are based on the assumption that the numerical coefficient introduced to correct for the approximations in the DCF method, $\fdcf$, is  0.5 \citep{ost01}. In the part of the simulated filament that we analyzed in detail, the field strength is intermediate between that of SR1 and of SR2. The measured value of $\fdcf$ was slightly larger than 0.5, but consistent with that value within the expected statistical uncertainties.

\item[4.] {\it Comparison with Zeeman field estimates.}
Based on the Zeeman data summarized by \citet{crut10}, \citet{li15} concluded that the average 3D magnetic field in molecular clumps in the ISM is $66n_{\htwom,4}^{0.65}$~\mug. For a typical inclination with respect to the plane of the sky of 30\degree, this corresponds to a POS field $B_0=57n_{\htwom,4}^{0.65}$~\mug. This is several times larger than the DCF estimate of the field in SR1, and we suggested that this could be due to the measured LOS velocity dispersion being less than the POS velocity dispersion. The POS field (for $\gamma=30^\circ$) corresponding to the average interstellar Zeeman field agrees reasonably well with the DCF field in SR2 and with the true POS field in the simulation. 

\item[5.] {\it Resolution effect on the magnetic field dispersion.}
The dispersion in polarization angles from the low resolution Planck data is significantly smaller than that of the high resolution HAWC+ data. 
\citet{hei01} and \citet{fal08} found this resolution effect in their simulations, and we do also. The simulation shows that the angle dispersion in low-column regions is less than in high-column regions, which contributes to the observed resolution effect.

\item[6.] {\it Polarization vectors and surface density gradients.}
The relative distribution of the inferred magnetic field vectors and the tangent of surface density contours in the observed B211 region shows that the magnetic field has a tendency to be normal to the contours of surface density. This can be quantified by the histogram shape parameter, $\xi$, defined in equation (\ref{eq:eta}). In B211, we find $\xi= -0.28$, meaning that the number of pixels with a magnetic field nearly normal to the contours of the surface density is about 1.8 times that with the magnetic field nearly parallel to the contours. The tendency for the field to be normal to the contours is primarily due to SR2, which has $\xi=-0.48$ and an average angle between the contours and the field of $\avg{\phi}=55^\circ$. The fact that there is some correlation between the orientation of the field and the column density contours of the gas indicates that the magnetic field is at least marginally dynamically important there. In the chaotic region SR1, the fact that $\xi=-0.03$ indicates that the magnetic field is dynamically sub-dominant, in agreement with the large value of the projected \alfven Mach number there (Table \ref{tab:properties}).

\item[7.] {\it Physical states of the two subregions and of the simulated filament.}
From the mass-to-flux ratios and \alfven Mach numbers, SR1 is magnetically supercritical and slightly super-\alfvenicstop, although we have suggested that the DCF method underestimates the field in SR1. The magnetic field in SR2 is significantly greater than that in SR1. Both the standard DCF analysis and the DCF/SF method suggest that SR2 is approximately magnetically critical and that it is trans-\alfvenicstop. The segment of the simulated filament we have analyzed is magnetically supercritical like SR1, although it has a significantly smaller dispersion of PA angles; it has an \alfven Mach number of about unity. The ratio of the line mass to the critical line mass is slightly less than unity for SR1, SR2, and the simulated filament if the full velocity widths of the filaments are used to estimate the virial parameters. Pre-stellar cores are suggested in both SR1 and SR2 \citep{mar16}.  There are two cores forming in the segment of the simulated filament that we have analyzed. 

\item[8.] {\it The DCF method.} In Appendix \ref{app:dcf}, we present derivations of both the standard DCF method and the structure function (SF) variant that are not restricted to small values of the polarization angles. We show that the standard DCF result often applies for the case of equipartition even if the perturbed field is not due to \alfven waves. Our simulation confirms that $\delta$, the correlation length of the turbulent magnetic field, is small, as assumed in the derivation of the DCF/SF method \citep{hil09}. For SR1, SR2 and our simulation, we find that $\delta\sim$~FWHM of the filament $\sim 0.1$~pc, consistent with the formation of a filament in a turbulent medium. We discuss the restriction procedure often used in the DCF/SF method in which differences in angles that exceed $90^\circ$ are converted to $|180-90^\circ|$ and suggest that restriction provides a lower limit on the structure function and is significant only when the dispersion in PAs is large, so that the DCF method is of questionable accuracy.

\item[9.] {\it Different versions of the DCF method.} In Appendix \ref{app:dcf} we also compare the standard DCF method with the DCF/SF version \citep{hil09} and the parallel-$\delta\vecB$ version \citep{ska21a,ska21b}. In most cases, the three methods agree within the uncertainties for both the observed regions, SR1 and SR2, and for the simulation. The exception is the standard DCF method, which gives a low value for the mean field in the highly tangled region SR1, probably because this method does not allow for spatial variation of $\vecB_0$.

\item[10.] {\it Equilibrium filaments and their fragmentation.} 
In Appendix B we give analytic estimates of the fragment mass
and the condition for the formation of a pre-stellar core in an unmagnetized filament.

\end{itemize}

\section*{Acknowledgments}
Support for this research is based on observations made with the NASA/DLR Stratospheric Observatory for Infrared Astronomy (SOFIA) under the 07\_0017 Program. SOFIA is jointly operated by the Universities Space Research Association, Inc. (USRA), under NASA contract NNA17BF53C, and the Deutsches SOFIA Institut (DSI) under DLR contract 50 OK 0901 to the University of Stuttgart. We thank Che-Yu Chen, Amitava Bhattacharjee, Martin Houde, Alex Lazarian, and Junhao Liu for a number of very helpful conversations. We also thank the two anonymous referees for their many helpful suggestions that greatly improve the paper. Support for this research was provided by NASA through a NASA ATP grant NNX17AK39G (RIK, CFM \& PSL) and the US Department of Energy at the Lawrence Livermore National Laboratory under contract DE-AC52-07NA 27344 (RIK). CFM acknowledges the hospitality of the Center for Computational Astronomy of the Flatiron Institute in New York, where he was a visiting scholar. JR thanks partial support from SOFIA program 07\_0017 and support from the SOFIA program 07\_0047 and NASA Astrophysics Data Analysis grant (80NSSC20K0449). PhA acknowledges support from “Ile de France” regional funding (DIM-ACAV+ Pro- gram) and from the French national programs of CNRS/INSU on stellar and ISM physics (PNPS and PCMI). This work is partly based on observations carried out under project number 129-15 with the IRAM 30m telescope. IRAM is supported by INSU/CNRS (France), MPG (Germany) and IGN (Spain). The present study also made use of data from the Herschel Gould Belt survey (HGBS) project (http://gouldbelt-herschel.cea.fr). This work used computing resources from an award from the Extreme Science and Engineering Discovery Environment (XSEDE), which is supported by the National Science Foundation grant number ACI-1548562, through the grant TG-MCA00N020, computing resources provided by an award from the NASA High-End Computing (HEC) Program through the NASA Advanced Supercomputing (NAS) Division at Ames Research Center, and an award of computing resources from the National Energy Research Scientific Computing Center (NERSC), a U.S. Department of Energy Office of Science User Facility located at Lawrence Berkeley National Laboratory,
operated under Contract No. DE-AC02-05CH11231.

\section*{Data Availability}
The processed HAWC+ data in FITS format is available at CDS via anonymous ftp to cdsarc.u-strasbg.fr (130.79.128.5) or via https://cdsarc.unistra.fr/viz-bin/cat/J/MNRAS.

\appendix{

\section{The Davis-Chandrasekhar-Fermi Method}
\label{app:dcf}

\citet{dav51} and \citet{cha53} proposed a method for estimating magnetic
field strengths in the ISM based on the assumptions that the medium is isotropic and that variations in the orientation of the field are due to \alfven waves. Hereafter, we refer to this as the DCF method. However, there are different approximations used and assumptions made in the literature, particularly in dealing with a large dispersion in the polarization angles (PAs). Therefore in this appendix we give a more rigorous derivation of the DCF result based on the method of Hildebrand et al. (2009). We then discuss two variants of the DCF method, the structure function method (DCF/SF) of \citet{hil09} and the parallel-$\delta B$ version of \citet{ska21a}. In applications of the DCF/SF method, differences between the PAs at different points are often restricted to be less than 90\degree, and we show how that can be problematic.

Only fields in the plane of the sky (POS) can be inferred in this manner, and, as noted in equation (\ref{eq:3d}), in this paper $\vecB$ (and $\vecv$) always refer to the components of the magnetic field and velocity in the POS. For \alfven waves, which are transverse, the equation of motion implies
\beq
\delta\vecv=\pm\frac{\delta\vecB}{(4\pi\rho)^{1/2}},
\label{eq:dv}
\eeq
where $\delta \vecv$ and $\delta\vecB$ represent the wave amplitude in the POS. For circularly polarized simple waves, this relation is valid for arbitrary wave amplitudes \citep{sher60}; for linearly polarized waves, it is valid only in the linear regime, since the wave is affected by the magnetic pressure gradients. This relation implies equipartition between the turbulent kinetic energy of motions normal to the mean magnetic field in the POS, $\vecB_0$, and the corresponding field energy in the waves, $\rho\delta v_\perp^2/2=\delta B_\perp^2/8\pi$, where the POS quantities $\delta \vecv_\perp$ and $\delta\vecB_\perp$ are perpendicular to the mean POS field. Under the assumption that the turbulent velocities are isotropic, the rms value of $\delta v_\perp$ is the same as the LOS velocity dispersion, $\sigma_V$. If the mean 3D field is at an angle $\gamma$ with respect to the POS and this angle is small enough that $\cos\gamma\simeq 1$, the assumption of isotropic velocities can be relaxed to be that the turbulent velocities are isotropic in the plane perpendicular to the mean 3D field. Let $\sdb$ be the rms value of $\delta B_\perp$. Equation (\ref{eq:dv}) then implies
\beq
\sigma_V\simeq\frac{\sdb}{(4\pi\rho)^{1/2}},
\label{eq:sigmav}
\eeq
where $\rho$ is a suitably averaged mean density. As discussed below, it is possible to measure the ratio $\sdb/B_0$. We can therefore obtain the value of $B_0$ by dividing both sides of this equation by $B_0$,
\beq
B_0\simeq  \frac{(4\pi\rho)^{1/2}\sigma_V}{\sdb/B_0}.
\label{eq:bo1}
\eeq

The same result holds under the more general assumption of equipartition of turbulent magnetic and kinetic energies in the POS, $\rho\delta v^2/2\simeq \delta B^2/8\pi$, provided that the fluctuations in the velocity and in the field are isotropic. Isotropy implies $\delta v^2=2\sigma_V^2$  and $\delta B^2 = 2\delta B_\perp^2$ (recall that $\delta v$ and $\delta B$ are POS quantities and thus two dimensional). Equipartition then implies $\rho\sigma_V^2=\sdb^2/4\pi$, which is the same as equation (\ref{eq:bo1}). \citet{hei01} found that the magnetic fluctuations were somewhat smaller than expected from equipartition, so that equation (\ref{eq:bo1}) overestimates $B_0$; this is taken care of by the factor $\fdcf$ in equation (\ref{eq:bo2.5}) below.

Isotropy is an important assumption in the DCF method. Observations by \citet{hey08} of the Taurus molecular cloud show that the turbulence there is anisotropic; it is not known if this is typical for molecular clouds. Their simulations for $\beta=2\cs^2/\va^2=0.02$ were strongly anisotropic, with 1D velocities perpendicular to the mean field 2-4 times greater than those along the field. As noted in the discussion of \alfvenic turbulence above, the DCF method can still be applied in the presence of such anisotropy if the mean 3D field is close to the POS. (The median value of the inclination $\gamma$ is 30\degree, for which $\cos\gamma\simeq 0.87\sim 1$.) For larger values of $\beta$, the simulations of \citet{hey08} for $\beta\geq 0.2$ and those of \citet{hei01} for $\beta\geq 0.05$ showed approximately isotropic turbulence.

Another important assumption that went into the derivation of equation (\ref{eq:sigmav}) for \alfven waves and equation (\ref{eq:bo1}) for the case of equipartition is that a single turbulent region dominates the signal along the LOS; if there is one dominant object along the LOS, its depth must be smaller than, or at most comparable to, the turbulent correlation length. If there are multiple turbulent regions, then $\sigma_V$ includes the differences in mean velocities of the regions and $\sdb$ includes the differences in the mean field orientation along the LOS. Such effects have been analyzed by \citet{zwe90}, \citet{myer91}, and \citet{hou09}. As discussed in Section \ref{sec:bobs}, possible effects of this type of inhomogeneity in the region we have observed are small.

The ratio $\sdb /B_0$ is estimated from fluctuations in the orientation of the field as revealed by polarization observations. We now discuss two methods of doing this, the standard method and the structure function method developed by \citet{hil09}. Bear in mind that a basic assumption of the DCF method is that the polarization traces an appropriately weighted (including by the density) integral of the direction of the magnetic field along the LOS. It must be borne in mind that the polarization angles (PAs), $\theta_i$, are limited to the range $-90^\circ\leq\theta_i\leq 90^\circ$, whereas the field angles (FAs), $\tfai$, extend over the range $-180^\circ\leq\tfai\leq 180^\circ$, so that that there is a 180\degree\ ambiguity in the relation between the FAs and the PAs. \citet{mar74} showed that the PA traces the mean FA in the simple case in which the FA is a linear function of position and the density is constant; for variable density, the conclusion holds if the FA is a linear function of surface density. 

\subsection{The Standard DCF method}
\label{app:standard}

Let the total POS field be $\vecB=\vecB_0+\delta\vecB$, where $\vecB_0=\avg{\vecB}$ is the mean POS field in the region being studied and $\avg{\delta\vecB}=0$. Let $\vecB_\parallel=\vecB_0+\delta\vecB_\parallel$ be the component of the POS field parallel to $\vecB_0$ and $\delta\vecB_\perp$ be the POS component perpendicular to $\vecB_0$. Then the field angle (FA), $\tfa$, at point $i$ relative to $\hat\vecB_0$ is:
\beq
\cos\tfai=\hat\vecB_i\cdot\hat\vecB_0=\frac{B_{\parallel,i}}
{(B_{\parallel,i}^2
+\delta B_{\perp,i}^2)^{1/2}}.
\label{eq:cosi0}
\eeq
This is presumably the density-weighted mean along the LOS for optically thin emission so that the PA coincides with the FA to within a 180\degree\ ambiguity. We now evaluate this under the assumption that $\delta B_i\ll B_0$ and then extend it to larger values. With this assumption, equation (\ref{eq:cosi0}) becomes 
\beq
\cos\tfai\simeq \frac{1}{\dis\left(1+\delta B_{\perp,i}^2/B_0^2\right)^{1/2}},
\eeq
with an error of order $\delta B_{\perp,i}^2\delta B_{\parallel,i}/B_0^3$.
The average value
of $\cos\tfai$ is then 
\beq
\avg{\cos\tfai}\equiv\cos\Delta\tfa=
\left\langle\frac{1}{\left(1+\delta B_{\perp,i}^2/B_0^2\right)^{1/2}}\right\rangle
\label{eq:cosi}
\eeq
with an error of order $\avg{\delta B_{\perp,i}^2\delta B_{\parallel,i}^2}/B_0^4$. Note that $\Delta\tfa$ depends only on perturbations perpendicular to the mean field; uniform compressions or rarefactions have no effect. Since the sign of $\tfa$ is irrelevant, we choose it to be positive. For a random field, $\avg{\cos\tfai}=0$ so that $\Delta\tfa=\pi/2$. Equation (\ref{eq:cosi}) shows that in this case $B_0=0$: Despite being derived under the assumption that $\delta B/B_0$ is small, this equation remains valid in the opposite limit. Note that while the average FA as measured by $\Delta\tfa$ must be in the range $0-\pi/2$, our analysis does not exclude the possibility that some individual FAs can exceed $\pi/2$. Defining $\sdb=\avg{\delta B_\perp^2}^{1/2}$, we approximate equation (\ref{eq:cosi}) as
\beq
\cos\Delta\tfa\simeq\frac{1}{(1+\sdb^2/B_0^2)^{1/2}}
\label{eq:cosdelta}
\eeq
with an error relative to that equation of order $(\sdb/B_0)^4$. Relating the cosine to the tangent, we then obtain
\beq
\tan\Delta\tfa\simeq \frac{\sdb}{B_0},
\label{eq:tan}
\eeq
so that (eq. \ref{eq:bo1})
\beq
B_0\simeq\frac{(4\pi\rho)^{1/2}\sigma_V}{\tan\Delta\tfa}.
\label{eq:bo2}
\eeq
Despite the approximations made, this result remains valid in the limit of a random field, for which $B_0=0$: In that case, $\avg{\cos\tfai}=0$ and $\Delta\tfa=\pi/2$ as noted above and $\tan\Delta\tfa=\infty$; equation (\ref{eq:bo2}) then gives $B_0=0$, as required.

We now make two approximations. First, to express the mean field in terms of the dispersion in the PAs, $\sigma_\theta=\avg{\theta_i^2}^{1/2}$, we note that the standard approximation $1-\cos\theta\simeq \frac 12\theta^2$ implies that $\Delta\theta\simeq\sigma_\theta$ from equation (\ref{eq:cosi}). This approximation for $1-\cos\theta$ is reasonably good even for relatively large values of $\theta$: For $\theta=\pi/2=1.57$, the approximation gives $\theta=[2(1-\cos\theta)]^{1/2}=\surd 2$, which is off by only 11 percent. The second approximation is central to the DCF method: We assume that for the most part the FAs are approximately equal to the PAs, so that $\cos\Delta\tfa\equiv\avg{\cos\tfai}\simeq\avg{\cos\theta_i}\equiv\cos\Delta\theta$. We combine these approximations to set
\beq
\tan\Delta\tfa\simeq\tan\sigma_\theta, 
\label{eq:tan2}
\eeq
which relates the average cosine of the FAs, which determines $B_0$, to the dispersion of the PAs, which is what can be observed.  For a Gaussian distribution of FAs, one can show that this approximation is accurate to within 10 percent for $\sigma_\theta<45^\circ$. The approximation is even more accurate for a uniform distribution of PAs (quite different from a Gaussian) with this dispersion; note that the distribution of FAs in the simulation of \citet{pado01} is much closer to a uniform distribution than to a Gaussian.\footnote{A uniform distribution of PAs changes significantly for $\sigma_\theta>52^\circ$. If the distribution of FAs extends over the range $\pm\theta_m$, then $\sigma_\tfa=\theta_m/\surd3$, so that a dispersion of 52\degree\ corresponds to $\theta_m=\pi/2$. For $\theta_m\leq\pi/2$, the PAs are identical to the FAs (to within an overall sign ambiguity of 180\degree). For $\sigma_\tfa$ between 90\degree\ and 180\degree, $\sigma_\theta$ is confined to the narrow range 52\degree-59\degree.} Equations (\ref{eq:tan}) and (\ref{eq:tan2}) then give the standard result for the dispersion of the component of the POS field perpendicular to the mean POS field,
\beq
\frac{\sdb}{B_0}\simeq \tan\sigma_\theta,
\label{eq:sdb0}
\eeq
although this is less accurate than equation (\ref{eq:tan2}). Correspondingly, the strength of the total POS field is $B=B_0\sec\sigma_\theta$. The resulting value of the mean POS field is then (eq. \ref{eq:bo2})
\beq
B_0=\fdcf \,\frac{(4\pi\rho)^{1/2}\sigma_V}{\tan\sigma_\theta},
\label{eq:bo2.5}
\eeq
where $\fdcf$ allows for inaccuracies in the approximations that led to this result. For $\fdcf=1$, the RHS of this equation is identical to the result of \citet{fal08}. The factor $\fdcf$ must be determined from simulations. Following \citet{ost01}, we set $\fdcf=0.5$ in this work. \citet{pado01} found $\fdcf=0.4\pm0.11$ in their analysis of the fields in three gravitationally bound cores in their simulation. In general, $\fdcf$ depends on the physical conditions and possibly on the resolution (Houde, private communication).

The accuracy of the DCF method depends upon both the dispersion of the PAs, $\sigma_\theta$, and on the angle between the mean field and the plane of the sky, $\gamma$, through its effect on $\sigma_\theta$. The method fails for $\gamma\simeq 90^\circ$, where $\sdb\gg B_0$ and $\tan\sigma_\theta$ becomes large. \citet{ost01} found that a sufficient condition for the DCF method to be accurate is $\sigma_\theta\leq 25^\circ$ and $\gamma\leq 60^\circ$. The approximations that led to equation (\ref{eq:tan}) become increasingly inaccurate as $\sdb/B_0$ increases, so it is best to keep $\sdb/B_0\simeq\tan\sigma_\theta<1$, corresponding to $\sigma_\theta <45^\circ$. A limit on $\sigma_\theta$ gives a limit on the 3D \alfven Mach number, $\ma$. The value of $\ma$ for the mean 3D field, $B_{0,\,\rm 3D}=B_0/\cos\gamma$, for isotropic turbulence is (eq. \ref{eq:bo2.5})
\beq
\ma=\frac{ \surd 3\tan\sigma_\theta\cos\gamma}{\fdcf}\rightarrow 3.5\tan\sigma_\theta\cos\gamma,
\label{eq:ma}
\eeq
where the last step is for $\fdcf=1/2$. For $\sigma_\theta<(25^\circ,\,45^\circ)$ this is $\ma<(1.6,\,3.5)\cos\gamma$.

Different assumptions lead to different approximations for $\sdb$. For example, \citet{zwe96} assumed that the FAs are identical to the PAs at the outset. She therefore excluded the possibility that individual FAs could exceed 90\degree, in contrast to our approach. With $\delta B_i=B_0\tan\theta_i$, averaging $\delta B^2$ over different lines of sight gives $\sdb^2/B_0^2=\avg{\tan^2\theta}$; she obtained the same result through an analysis using the Stokes parameters. \citet{hei01} recognized that this is problematic for flows with \alfven Mach numbers $\ga 1$ since the average of $\tan^2\theta$ is dominated by angles near $90^\circ$, and they suggested several approximations to overcome this. As noted above, \citet{fal08} suggested replacing the average of the tangent by the tangent of the average, which we derived above; this also overcomes this problem.

\subsection{The Structure Function Version of the DCF Method (DCF/SF)}
\label{app:structure}

\citet{hil09} improved on the standard DCF approach by allowing the direction of the mean magnetic field to be a slowly varying function of position, $\boox$; the magnitude of the field was assumed to be constant, however.  A strength of their method is that the unknown direction of the mean field is not needed in order to determine its magnitude. Furthermore, it is relatively independent of the dispersion of PAs on large scales and can therefore handle cases in which large dispersions on large scales cause the standard DCF method to break down. \citet{hou09} extended this method to allow for variations along the line of sight and across the telescope beam, but at the expense of adding an additional parameter that must be fit from the data. Here we follow the simpler approach of \citet{hil09}. We include possible effects of integration along the line of sight, in addition to the effects of other approximations made in the method, in a numerical factor $\fdcf$, as in equation (\ref{eq:bo2}).

The field is decomposed into
a smooth part and a turbulent part, 
\beq
\bx=\boox+\delta\bx
\eeq
with 
\beqa
\avg{\delta\bx}&=&0,\\ 
\avg{\delta\bx\cdot\vecB_0(\vecx+\vecl)}&=&0,
\eeqa
where the average is taken over the observed area and $\vecl$ is constant. Note that since $\boox$ is now a function of position, the value of $\sdb$ differs from that in the standard method, in which $\boox$ is assumed to be constant. They then evaluate the two-point correlation function\footnote{Actually, they defined the correlation  $\avg{\cos\Delta\Psi(\ell)}$ as $\avg{\vecB(\vecx)\cdot\vecB(\vecx+\vecl)}/\avg{B^2(\vecx)}^{1/2}\avg{B^2(\vecx+\vecl)}^{1/2}$, which agrees with the exact expression if $B(\vecx)B(\vecx+\vecl)$ is uncorrelated with $\cos\Delta\Psi(\ell)$ and if $\avg{B(\vecx)B(\vecx+\vecl)}=\avg{B^2(\vecx)}^{1/2}\avg{B^2(\vecx+\vecl)}^{1/2}$. In the end, their approximations and ours lead to the same result.}
\beq
\avg{\cos\Delta\Psi(\ell)}=\left\langle\hat\vecB(\vecx)\cdot\hat\vecB(\vecx+\vecl)\right\rangle.
\eeq
Making the approximation that the average of the ratio is the ratio of the averages yields
\beq
\avg{\cos\Delta\Psi(\ell)}=\frac{\avg{\boox\cdot\booxl}+\avg{\delta\bx\cdot\delta\bxl}}{\avg{B^2}}.
\label{eq:avgsf}
\eeq
Since 
\beq
\avg{B^2}=B_\rms^2= B_0^2+\sigma_{\delta B}^2, 
\label{eq:brms}
\eeq
this approximation has eliminated the effect of non-zero values of $\delta\vecB\cdot\vecB_0$ on the analysis. We then have
\beqa
1-\avg{\cos\Delta\Psi(\ell)}&=&\\
&&\hspace{-3cm}\frac{[B_0^2-\avg{\boox\cdot\booxl}]+[\sigma_{\delta B}^2-\avg{\delta\bx\cdot\delta\bxl}]}{B_\rms^2}.\nonumber
\label{eq:omcos}
\eeqa
\citet{hil09} assumed that $\booxl$ is slowly varying and expanded it in powers of $\ell$.  The linear term averages out, so the lowest order term varies as $\ell^2$. \citet{hil09} made the small angle approximation, but we follow \citet{hou09} in not doing that yet. This equation then becomes
\beqa
\hspace{-0.7cm} 1-\avg{\cos\Delta\Psi(\ell)}&=&
\label{eq:one}\\
&&  \hspace{-1cm} \frac 12 m^2\ell^2+\frac{\sigma_{\delta B}^2}{B_\rms^2}-\frac{\avg{\delta\bx\cdot\delta\bxl}}{B_\rms^2},\nonumber
\eeqa
where $m$ is a constant that is determined by fitting the data. Note that this equation is the same as would have been obtained had we assumed that $\delta\vecB\cdot\vecB_0=0$, a result of the approximation made in equation (\ref{eq:avgsf}).

\citet{hil09} further assumed that the last term vanishes for length scales $\ell$ exceeding the correlation length of the turbulent field, $\delta$. Let $1-\cos\Delta\Psi_0$ be the value of $1-\avg{\cos\Delta\Psi(\ell)}$ obtained by extrapolating the first two terms of  this equation from large $\ell$, where the last term is negligible, to $\ell=0$:
\beq
1-\cos\Delta\Psi_0= \frac{\sigma_{\delta B}^2}{B_\rms^2},
\label{eq:delta}
\eeq
so that with the aid of equation (\ref{eq:brms})
\beq
\frac{\sigma_{\delta B}}{B_0}=\left(\frac{1}{\cos\Delta\Psi_0}-1\right)^{1/2}.
\label{eq:sdb1}
\eeq
The total POS field strength is $B=B_0/(\cos\Delta\Psi_0)^{1/2}$. The result for $B_0$ is then
\beq
B_0=(4\pi\rho)^{1/2}\sigma_V\,\left(\frac{1}{\cos\Delta\Psi_0}-1\right)^{-1/2}
\label{eq:bo3}
\eeq
from equation (\ref{eq:bo1}).
Note that in contrast to our derivation of the standard DCF method, it is the total dispersion in the POS field, $\sigma_{\delta B}$, that enters equation (\ref{eq:sdb1}) rather than the dispersion perpendicular to the mean field, $\sdb$ (eq. \ref{eq:cosdelta}). They assumed that $\delta\vecB$ is isotropic, and in that case the difference is small.

As for the standard DCF method, one must then assume that the FAs are approximately equal to the PAs. Following \citet{hil09} we label the PAs by $\Phi$. The value of $\Phi$ is the same as that of $\theta$ in section \ref{app:standard} if angles are measured relative to the mean field direction. The value of $B_0$ in terms of measurable quantities is then
\beq
B_0=\fdcf(4\pi\rho)^{1/2}\sigma_V\,\left(\frac{1}{\cos\Delta\Phi_0}-1\right)^{-1/2},
\label{eq:bo3.5}
\eeq
where $\fdcf$ allows for inaccuracies in the approximations that led to this result.  \citet{hil09} did not include such a factor. As noted above, \citet{hou09} explicitly allowed for variations along the line of sight, but did not correct for the effect of the approximations in the method. In the text, we set $\fdcf=0.5$. The 3D \alfven Mach number with respect to the mean 3D field is
\beq
\ma=\frac{\surd 3}{\fdcf}\left(\frac{1}{\cos\Delta\Phi_0}-1\right)^{1/2}\cos\gamma.
\label{eq:ma2}
\eeq
For small values of $\Delta\Phi_0$, this reduces to $\ma\simeq (\surd 3\cos\gamma/\fdcf)\Delta\Phi_0/\surd 2$, which agrees with the result for the standard DCF method (eq. \ref{eq:ma}) for small $\sigma_\theta$ if $\Delta\Phi_0/\surd 2$ is replaced by $\sigma_\theta$ (see below eq. \ref{eq:bo4}).

\citet{hil09} made the small-angle approximation, retaining terms of order $\Delta\Phi_0^2$, and assumed
$\Delta\Psi_0=\Delta\Phi_0$, so that equation (\ref{eq:sdb1}) becomes
\beq
\frac{\sigma_{\delta B}}{B_0}=\frac{\Delta\Phi_0}{(2-\Delta\Phi_0^2)^{1/2}},
\label{eq:sdb2}
\eeq
where $\Delta\Phi_0$ is inferred from $\Delta\Phi(\ell)$ in the same way that $\Delta\Psi_0$ is inferred from $\Delta\Psi(\ell)$ as described above. This approximation is accurate to within 10 percent for $\Delta\Phi_0<60^\circ$. For small angles, $\Delta\Phi(\ell)$ is given by equation (\ref{eq:sf0}).
Equation (\ref{eq:bo3.5}) for the mean magnetic field becomes
\beq
B_0=\fdcf(4\pi\rho)^{1/2}\sigma_V\,\frac{(2-\Delta\Phi_0^2)^{1/2}}{\Delta\Phi_0}.
\label{eq:bo4}
\eeq
For $\fdcf=1$, this agrees with their result since their $b$ is $\Delta\Phi_0$. The factor $(2-\Delta\Phi_0^2)^{1/2}$ arises because equation (\ref{eq:delta}) gives $\Delta\Phi_0$ in terms of $\avg{B}$ instead of $B_0$. For small $\Delta\Phi$, this result agrees with equation (\ref{eq:bo2}) for the standard DCF method: $\Delta\Phi_0$ is an average of the difference of two random angles, so that $\Delta\Phi_0=\surd 2 \sigma_\theta$ \citep{hil09}. The agreement of the two expressions for $B_0$ implies that the value of the correction factor $\fdcf$ is the same for the two methods. 

\citet{hil09} assumed that $\delta$, the maximum scale of the turbulent velocity correlations, was of order 1 mpc, well below the resolution of the observations they were fitting. Subsequent work using the method of \citet{hou09} obtained larger values: for example, \citet{gue21} found $\delta\sim 10-100$ mpc for OMC-1. Our analysis of SR1 and SR2 also gives $\delta\sim 100$~mpc, as does our simulation (see below). The observations we have analyzed and our simulation are consistent with the turbulent correlation length $\delta$ being of order the FWHM of the filament, which is plausible for a filament formed in a turbulent medium and consistent with the results of \citet{pal13}. Note that the value of $\delta$ does not enter; all that is required is that it be small enough that there is a range of $\ell$ over which $\Delta\Phi(\ell)^2$ is accurately fit by the first two terms in equation (\ref{eq:one}).

How well does our simulation agree with the SF variant of the DCF method? The actual values of the field strength are compared with the DCF/SF values in Table \ref{tab:dcf_sf_sim}. Here we focus on the validity of the SF relation between the dispersion in field angles and the dispersion in field strength, equation (\ref{eq:one}). For simplicity, we adopt the small angle approximation; had we not done that, the results would have changed by only 7\%. Fig. \ref{fig:A21} compares the dispersion in angle, $\Delta\Psi(\ell)$, measured in the FOV window of the simulation with values from  equation (\ref{eq:one}). The blue curve plots the first two terms in the equation, using the measured value of $\sdb/B_\rms$ and treating $m$ as a free parameter. It provides an excellent fit to the data for $\ell\ga 5$ superpixels, or about 0.1 pc for the parameters we have adopted. This is the turbulent correlation length, $\delta$, and is about 1/3 the width of the filament, as noted above. It is well resolved in the simulation, with more than 40 grid cells at the highest resolution. The red curve shows the last term in equation (\ref{eq:one}). As assumed by \citet{hil09}, it is negligible except at small scales, $\ell<\delta\sim 0.1$~pc. The figure shows that the approximations made in deriving equation (\ref{eq:one}) are reasonably good in this case, and that the turbulent velocity correlations extend to scales large enough that they must be taken into account.

\begin{figure}
\includegraphics[angle=0,scale=0.33]{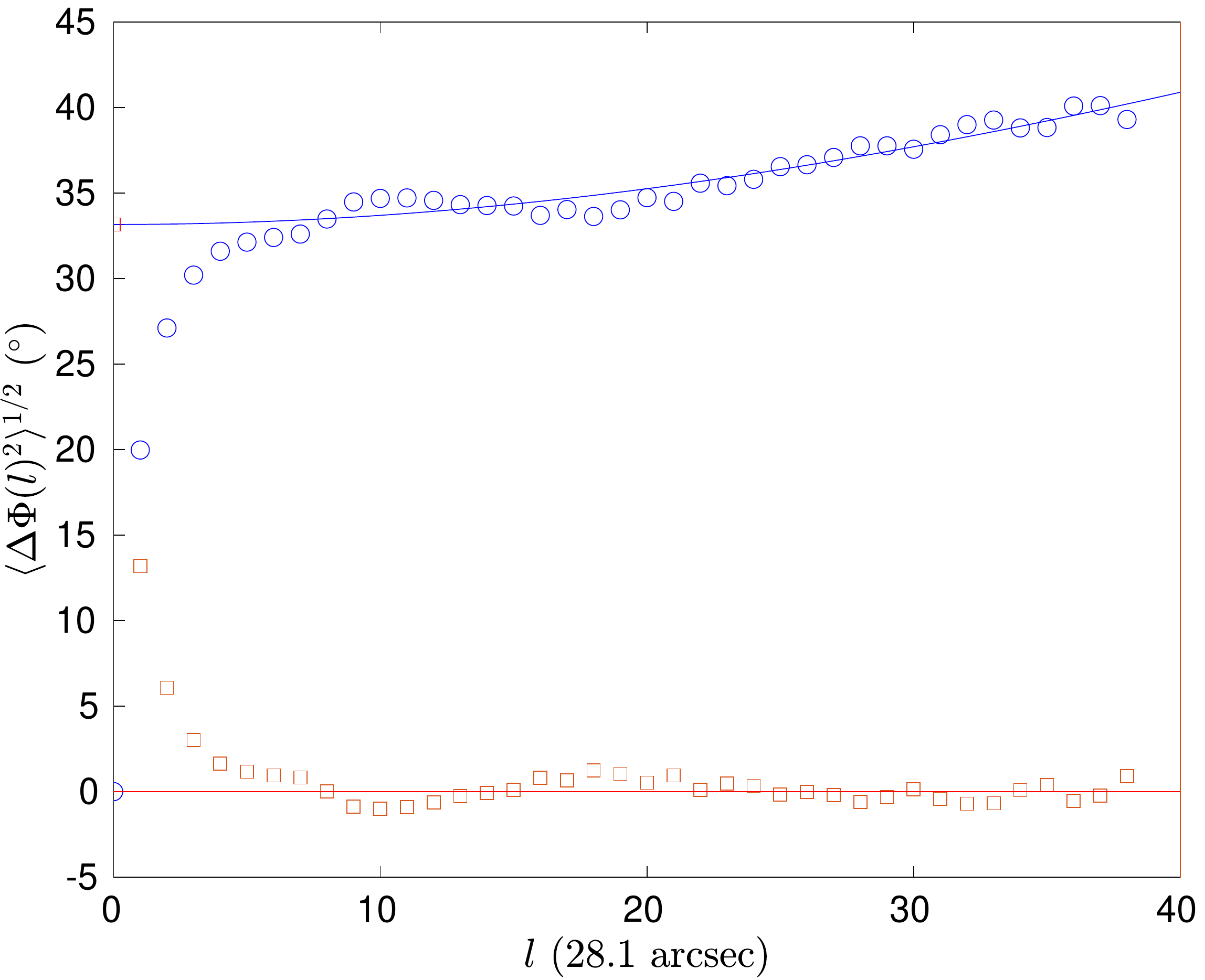}
\caption{The structure function of magnetic field vectors in the simulation window (blue circles) as a function of scale in units of HAWC+ superpixels (28.1 arcsec), with the best fit (solid curve) and the negative of the last term of equation (\ref{eq:one}) (red squares).
\label{fig:A21}}
\end{figure}

\subsubsection{Restriction of $\Delta\Phi$}
\label{app:restrict}

A fundamental problem with determining the field strength from polarization observations is that the field angles (FAs), $\Psi$,  can range over 360 degrees whereas the PAs, $\Phi$, are limited to a range of 180 degrees. For a given choice of the direction corresponding to 0\degree, let the orientation of the PAs lie in the range $\pm90$\degree. FAs lying outside that range will have PAs in the opposite direction--i.e., such an FA will differ from the corresponding PA by 180\degree. As a result, the measured value of $\Delta\Phi$, based on the PAs, will differ from the actual value, which is based on the FAs. The error depends on the number of FAs that are flipped in direction, which in turn depends on the choice of the 0\degree\ direction; we choose that to be the direction that gives the minimum dispersion in the PA angles, $\Phi_i$. In an attempt to reduce this error in the DCF/SF method, it is common to restrict the difference between angles, $\Delta \Phi$, to be less than 90\degree\ by replacing $|\Delta\Phi|$ with $|180^\circ-\Delta\Phi|$ when $|\Delta\Phi|>90$\degree\ (e.g., \citealp{dav11}). Under what conditions is this valid? If the angles are restricted when they should not be, then the dispersion will be underestimated and the field overestimated.

First assume that the PAs are an accurate reflection of the FAs, up to an ambiguity of 180\degree. If the FAs are confined to a range less than 180\degree, then the FAs and PAs can be in alignment and restricting $\Delta\Phi$ would lead to an error.  If the FAs extend beyond that range, but the mean field has a constant direction, then the dispersion in the PAs will be less than that in the FAs and the the field strength will be overestimated. The error will only grow larger if $\Delta\Phi$ is restricted. It follows that restriction should never be used if the mean field has a constant direction. 

If $\vecB_0$ changes direction as a function of position, the situation becomes more complicated. If the average angle differs significantly from the FAs in a local region, then it is possible that some of those FAs will be flipped by 180\degree\ when converted to PAs, thereby increasing the dispersion relative to neighboring PAs. Restriction corrects this by significantly reducing $\Delta\Phi$ for the PAs in that region. On the other hand, it also reduces $\Delta\Phi$ for the PAs that were initially quite different from the initial FA direction. This discussion suggests that restriction provides a lower limit on $\Delta\Phi_0$ and that it is significant only when the dispersion in angles is large, when the DCF method is of questionable accuracy. For SR2, with $\sigma_\theta=20^\circ$, restriction reduced $\Delta\Phi_0$ by 4\% and therefore increased $B_0$ by the same factor; for the simulation, with $\sigma_\theta=29^\circ$, restriction increased the inferred field by 5\%; and for SR1, with $\sigma_\theta=54^\circ$, restriction increased $B_0$ by 20\%. An approach that reduces the uncertainties due to restriction is to map the field locally \citep{gue21}, so that there is less variation of the mean field in each region.

\subsection{The parallel-\texorpdfstring{$\delta \vecB$}~ version of the DCF method}

\citet{ska21a} and \citet{ska21b} adopt an alternative approach to inferring the mean field strength and assume that the turbulent motions are in approximate equipartition with the {\it parallel} component of the perturbed field, $0.5\rho\delta v^2=B_0\delta B/4\pi$ for small $\delta B$. Setting $\sigma_\theta=\delta B/B_0$, they obtained
\beq
B_{0,\rm ST}=(2\pi\rho)^{1/2}\,\frac{\sigma_V}{\sigma_\theta^{1/2}}.
\eeq
They did not find it necessary to introduce a correction factor $\fdcf$ as is often done for the standard DCF method. They present the results of simulations showing that their result is more accurate than the standard one. 

In our simulation, we find that $\avg{|\delta\hat\vecB\cdot\hat\vecB_0|}\simeq 0.8$, so the parallel component of $\delta\vecB$ is indeed significant. On the other hand, the positive and negative values nearly cancel so that $\avg{\delta\hat\vecB\cdot\hat\vecB_0}\simeq 0.06$--i.e., rarefactions, which have $\delta\hat\vecB\cdot\hat\vecB_0<0$, nearly cancel the effect of compressions, which have $\delta\hat\vecB\cdot\hat\vecB_0>0$. This effect is not included in the model of \citet{ska21a} since they assumed that $\avg{\delta\vecB\cdot\vecB_0}$ can vanish only for incompressible turbulence and proceeded to make the incorrect assumption that $\avg{\delta\vecB\cdot\vecB_0}=\avg{|\delta \vecB|} B_0$. They attempted to justify this step by appealing to \citet{bha98}, although that work applies only to very subsonic turbulence and has $\avg{\delta\vecB}=0$. \citet{ska21b} argued that the maximum kinetic energy in fluctuations, a second order quantity, is in equipartition with the maximum magnetic energy in the fluctuations, a first order quantity; as shown by \citet{zwe95}, however, it is the second order energies that are in equipartition. In agreement with equation (\ref{eq:sdb0}), they note that a non-zero polarization angle is possible only in the presence of a perpendicular component of the field. As a result their method requires $\delta B_\parallel\simeq\delta B_\perp$, which they find to be satisfied to within a factor 2 in the simulations they analyze.

For $\tan\sigma_\theta\simeq \sigma_\theta$, the ratio of their result to the standard DCF one is $(\sigma_\theta/2)^{1/2}/\fdcf$. For $\fdcf=0.5$, the two values of the field agree for $\sigma_\theta=27^\circ$; since that is close to the values we find in our simulation, we are not able to determine whether their result is more accurate than the standard DCF method. It should be noted, however, that their result has no free parameters, whereas $\fdcf$ is a free parameter for the standard method. In view of the questionable assumptions underlying their method, more work is needed to understand the physical basis for the method and the circumstances under which it works.

Applying their method to the observed fields in B211 with the full line width $\sigma_V^m$ gives $B_0=27$~\mug\ for SR1, about twice the value with the standard DCF method but only slightly larger than the 23~\mug\ with the DCF/SF method with restriction. If $\sigma_\theta$ is replaced by $\tan\sigma_\theta$ in their formula, their result would be 22~\mug\ for this region, not that much larger than the DCF/SF value without restriction, 16~\mug. For SR2 they find $B_0=57$~\mug, somewhat less than the 66~\mug\ with the standard DCF method and $\sim 80$~\mug\ with the DCF/SF method. In most cases the three methods agree within the uncertainties for both observation and simulation. The exception is the standard DCF method, which gives a low value for the tangled-field SR1, most likely because it includes variation in $\vecB_0$ in its determination of $\sigma_\theta$. The parallel-$\delta\vecB$ method includes such variation as well, but the result is less sensitive since it enters only as the square root.

\section{Equilibrium and Fragmentation of Filaments}
\label{app:equil}

\subsection{Equilibria of Cylindrical Filaments}

Under what conditions are the filaments that we have observed and simulated expected to be stable against gravitational collapse?
 \citet{fie00a} have shown that the
maximum mass per unit length of an unmagnetized, equilibrium filamentary cloud is $M_{\vir,\ell}=2\avg{\sigma_V^2}/G$.
The virial parameter for a filament, $\avir$, is the ratio of twice the 2D kinetic energy to the magnitude of the potential energy and is given by
\beq
\avir=\frac{M_{\vir,\ell}}{M_\ell}=\frac{2\sigma_V^2}{GM_\ell}.
\label{eq:avir}
\eeq
Equilibria require $\avir>1$. In contrast to the spherical case, the gravitational energy term in the virial theorem is independent of the internal structure of the filament, so long as the density is independent of azimuth and distance along the filament.

The stability of a cloud against gravitational collapse is also affected by magnetic fields, which are parameterized by the mass-to-flux ratio relative to the critical value, $\muphi$. Let $M_\Phi$, the magnetic critical mass, be the maximum mass that can be supported by magnetic fields against gravity; then $\muphi=M/M_\Phi$. In general,
\beq
M_\Phi=\frac{c_\Phi\Phi}{G^{1/2}},
\label{eq:mphi}
\eeq
where $\Phi$ is the magnetic flux threading the cloud and $c_\Phi=1/(2\pi)\simeq 0.16$ for a thin disk \citep{naka78} and 0.17 for a spheroidal cloud with a constant mass-to-flux ratio \citep{tomi88}. The field in the ambient cloud is generally perpendicular to the filament when self-gravity is important (e.g., \citealp{pla16}); the filament can then grow by flows along the field lines \citep{pal13}. We shall focus on the case of a perpendicular field here; \citet{nag87,fie00a} and \citet{moti21} have considered the case in which the field is parallel to the filament. We anticipate that the critical mass per unit length of a filament is obtained from equation (\ref{eq:mphi}) by dividing both sides by the length, and indeed that is what \citet{tomi14} found for the case of a filament with a mass-to-flux distribution corresponding to a constant-density filament threaded by a uniform field. \citet{kash21} generalized this analysis to polytropic filaments. For $\gamma\rightarrow 1$, where here $\gamma$ is the adiabatic index, their result is within 1\% of the result expected from the case of a thin disk,
\beq
\mpl= \frac{\Phi_\ell}{2\pi G^{1/2}},
\label{eq:mpl}
\eeq
where $\Phi_\ell=w\botd$ is the flux per unit length, $\botd$ is the mean 3D field in the filament, and $w$ is the width of the filament. (They defined $\Phi_\ell$ as half the flux per unit length, so their coefficient is twice as large.) Since the mean surface density is $\Sigma=M_\ell/w$, the mass-to-flux ratio relative to the critical value is
\beq
\muphi=\frac{M_\ell}{\mpl}=\frac{2\pi G^{1/2}\Sigma}{\botd}.
\label{eq:muphi}
\eeq

The derivation of the magnetic critical mass neglects the presence of turbulent magnetic fields. Since it is the total field energy that counteracts the effect of gravity, we assume that it is the total 3D field, $\btd=(\botd^2+\delta \btd^2)^{1/2}$, that enters equation (\ref{eq:muphi}). The value of $\muphi$ that we can measure depends on the POS values of the field and of the surface density (we have added the subscript ``POS" to the total POS field for clarity):
\beq
\muphipos=7.6\times 10^{-21}N(\htwom)_{\rm Obs}/B_{\POS}, 
\label{eq:muphipos}
\eeq
where $B_\POS$ is measured in \mug\ and $N(\htwom)_{\rm Obs}$ is the observed column density of the filament. The stability of the filament depends on the column density normal to the filament, which is smaller than that by $\cos\gamma_f$, where $\gamma_f$ is the inclination angle of the filament relative to the POS. The actual value of $\muphi$ is then related to $\muphipos$ by
\beq
\muphi=\muphipos\cos\gamma_f\left(\frac{B_{\POS}}{\btd}\right).
\label{eq:cositot}
\eeq
We note that \citet{li15} showed that the volume-averaged field, which enters $\muphi$, is generally less than the mass-averaged field determined from Zeeman observations. If this same effect occurs for DCF determinations of the field, which are also mass-averaged, then the observed value of $\muphipos$ is an underestimate of the true value.

When the filamentary cloud is supported by both a perpendicular magnetic field and thermal/turbulent motions, \citet{kash21} found that the maximum stable mass per unit length for $\gamma\rightarrow 1$ is
\beq
M_{\crit,\ell}\simeq \left(M_{\Phi,\ell}^2+M_{\vir,\ell}^2\right)^{1/2},
\label{eq:crit}
\eeq
with a factor 0.85 before $M_{\vir,\ell}^2$; we have omitted that factor in order to make the result exact in the limit $\Phi_\ell=0$. Equation (\ref{eq:crit}) implies
\beq
\frac{M_\ell}{M_{\crit,\ell}}=\frac{1}{\left(\muphi^{-2}+\avir^2\right)^{1/2}}.
\label{eq:mellcr}
\eeq
Equilibrium clouds must have $\muphi^{-2}+\avir^2>1$ so that $M_\ell$ is less than the critical value.

\subsection{Fragmentation of Filaments Stable Against Radial Collapse}
\label{app:frag}

In the text, we find that the filaments SR1 and SR2 have $\avir>1$, so they are stable against radial collapse. Can they fragment? We begin with the case $B=0$ since the effects of magnetic fields have been considered only for fields parallel to the filament. Self-gravitating, isothermal filaments are characterized by the ratio of the radius to the scale height, $H=\cs/(4\pi G\rho_c)^{1/2}$, where $\rho_c$ is the central density. With the aid of the results of \citet{fis12}, this ratio is
\beq
\frac{R}{H}=\left(\frac{8}{\avir-1}\right)^{1/2}.
\label{eq:rh}
\eeq
Note that \citealp{fis12} express their results in terms of $f_{\rm cyl}=M_\ell/M_{\vir,\ell}=1/\avir$. \citet{nag87} studied the stability of isothermal filaments and found two types of behavior. For large $R/H$, gas compresses along the filament with a maximum growth rate at a wavenumber $k_m =0.284/H$ (the ``compressible instability"). For small $R/H$, the gas flow is almost incompressible (the ``deformation instability"), with a maximum growth rate at $k_m=0.58/R$. Combining these results, we obtain the approximation
\beqa
k_m H & \simeq &\left[0.284^2+\frac{0.58^2}{(R/H)^2}
\right]^{1/2},\\
&\simeq &0.204(\avir+1)^{1/2},
\label{eq:km}
\eeqa
where the second expression was obtained with the aid of equation (\ref{eq:rh}).
The latter expression agrees with the results of \citet{nag87} to within a few percent. A more accurate approximation, a fourth-order polynomial, is given by \citet{fis12}. It is convenient to express this in terms of the FWHM of the filament, $F$, which is observable. The results of \citet{fis12} can be fit to within 10\% by the expression $F/H=4.9/(\avir+0.28)^{1/2}$ so that
\beq
k_mF\simeq 1.00\left(\frac{\avir+1}{\avir+0.28}\right)^{1/2},
\eeq
which is within about 10\% of the curve for $\lambda_m$ in Fig. 11 of \citet{fis12}. Observe that the ratio of the wavelength of the fastest growing mode to the FWHM of the filament is almost constant, with $k_mF$ varying from 1.25 to 1.00 as $\avir$ increases from 1 to $\infty$.

The growth rate of the instability is within 25\% of $0.3(4\pi G\rho_c)^{-1/2}$ for all $R/H$ based on \citet{fis12}'s fit to the results of \citet{nag87}. Thus, unmagnetized filaments that are stable against radial collapse always fragment, although the amplitude of the perturbation could be small, as we shall now see.

The fragment mass is 
\beq
M_{\rm frag}=M_\ell\lambda=\frac{4\pi\cs^2}{\avir Gk},
\eeq
where $\lambda$ is the wavelength of the instability. An isolated fragment will settle into equilibrium for $M<\mbe$, where $\mbe=1.182\cs^3/(G^3\rho_s)^{1/2}$, the critical Bonnor-Ebert mass, is the maximum equilbrium mass for an isothermal sphere \citep{inu97}. The ratio of the fragment mass to the Bonnor-Ebert mass is
\beq
\frac{M_{\rm frag}}{\mbe}=\frac{(4\pi)^{1/2}}{1.182}\left(\frac{\avir-1}{\avir^2 kH}\right),
\label{eq:fbe0}
\eeq
which is equivalent to the result of \citet{fis12}. In deriving this expression, one must keep in mind that $H$ is defined in terms of the central density whereas $\mbe$ is defined in terms of the density at the surface. We expect that the wavenumber $k$ corresponds to the fastest growing mode, so equation (\ref{eq:km}) implies
\beq
\frac{M_{\rm frag}}{\mbe}=14.7\,\frac{(\avir-1)}{\avir^2(\avir+1)^{1/2}}.
\label{eq:fbe}
\eeq
For $\avir\gg1$, this result shows that $M_{\rm frag}\ll \mbe$; self-gravity is not important and the density in the fragment is not much larger than the mean density in the filament. The fragment mass exceeds $\mbe$ for $1.12<\avir<4.8$ ($0.89>f_{\rm cyl}>0.21$), and an isolated unmagnetized fragment would be expected to undergo gravitational collapse under these conditions. The optimum condition for fragmentation occurs when this ratio is a maximum, at $\avir\sim 2$ ($f_{\rm cyl}\sim 0.5$). These results are consistent with the graphical results in Figure 11 of \citet{fis12}. More generally, in the absence of magnetic fields pre-stellar cores (i.e., starless cores with $M\geq\mbe$) would be expected to form for $1.1\la\avir\la 5$ ($0.9\ga f_{\rm cyl}\ga 0.2$), although magnetic fields can inhibit their formation and collapse (see below). In their SPH simulations, \citet{inu97} found that a filament with $f_{\rm cyl}=0.2$ produced a stable core, consistent with our expectation; however, filaments with $f_{\rm cyl}=0.9$ produced cores that collapsed, contrary to expectation from equation (\ref{eq:fbe}). Their simulation had periodic boundary conditions, so the results are not expected to be identical to those for an isolated filament. They found that stable fragments could merge; gravitational collapse would ensue if the mass of the merged fragments exceeded $\mbe$.

\citet{moti21} studied the stability of polytropic filaments with a field parallel to the filament. For $B=0$ and $\gamma=\frac 34$ (the case closest to $\gamma=1$), they found that the most unstable wavelength and the critical wavelength are within about 30\% of the values for an isothermal filament (eq. \ref{eq:km}). This suggests that there is no significant difference between the unstable wavelengths of isothermal filaments and polytropic filaments with $\gamma\rightarrow 1$ despite the fact that isothermal filaments have a much steeper density gradient at large radii, $\rho\propto r^{-4}$ (e.g., \citealp{ostr64}), than polytropic filaments, $\rho\propto r^{-2/(2-\gamma)}$ \citep{via74}. The results discussed above should therefore apply to both isothermal filaments and polytropic ones with $\gamma\rightarrow 1$. We note that observationally, the polytropic solution appears to be favored -- for example, \citet{pal13} found $\rho\propto r^{-2}$ with $\gamma$ just below 1.

Simulations have shown that magnetic fields reduce fragmentation (e.g., \citealp{henn11,myer13}). A magnetic field parallel to the filament reduces the growth rate of the fragmentation instability, particularly for large $\avir$ (small $f_{\rm cyl}$), but it does not prevent instability \citep{nag87}. Fragmentation in the presence of a perpendicular field has not been analyzed to our knowledge, but we anticipate that fields with $\muphi<1$ (magnetically subcritical) would be stable. Subcritical magnetic fields are rarely observed, however \citep{crut10}. Although a strong perpendicular magnetic field is required to suppress fragmentation, a weaker field can prevent gravitational collapse of an approximately spherical fragment: For such a clump, the critical mass is $M_{\rm crit}\simeq \mbe+M_\Phi$ \citep{mcke89}: Both kinetic and magnetic energies contribute to stability, as in the case of stability against radial collapse. For a fragment, the relation analogous to equation (\ref{eq:mellcr}) for filaments is
\beq
\frac{M_{\rm frag}}{M_{\rm crit}}=\frac{1}{\dis\frac{\mbe}{M_{\rm frag}}+\muphi^{-1}},
\label{eq:cf}
\eeq
where the first term is given in equation (\ref{eq:fbe}). Fragments with $M_{\rm frag}>M_\crit$ are expected to undergo gravitational collapse.

\label{lastpage}

\end{document}